\documentclass[preprint, review,12pt] {elsarticle}

\usepackage{graphics}
\usepackage{graphicx}
\usepackage{epsfig}

\usepackage{amssymb}
\usepackage{amsthm}

\usepackage{lineno}
\usepackage{color}

\begin{document}

\begin{frontmatter}

 

\title{
%
From Bifurcations to State-Variable Statistics in Isotropic Turbulence: Internal Structure, Intermittency, and Kolmogorov Scaling via Non-Observable Quasi-PDFs
}
 

\author{Nicola de Divitiis}

\address{"La Sapienza" University, Dipartimento di Ingegneria Meccanica e 
Aerospaziale, Via Eudossiana, 18, 00184 Rome, Italy, \\
phone: +39--0644585268, \ \ fax: +39--0644585750, \\ 
e-mail: n.dedivitiis@gmail.com, \ \  nicola.dedivitiis@uniroma1.it}

\begin{abstract} 
This article investigates the intrinsic link between skewness and statistical intermittency in velocity and temperature increments within homogeneous isotropic turbulence. The theoretical framework builds upon the author’s previously established closure schemes for the von K\'arm\'an--Howarth and Corrsin equations. A transition Taylor--scale Reynolds number is first estimated via a formal bifurcation analysis of the closed von K\'arm\'an--Howarth equation. 

A central thesis of this work is that while the nonlinearity of the Navier--Stokes equations is fundamentally responsible for the emergence of intermittency in velocity and temperature increments, it is insufficient on its own to recover the Kolmogorov scaling law. We demonstrate that the non--observability of bifurcation modes constitutes the missing conceptual link: the concomitant effect of nonlinearity and non--observability not only determines the Kolmogorov scaling and drives an intermittency that grows monotonically with the Taylor--scale Reynolds number, but also enables the analytical determination of the internal structure functions of velocity and temperature differences, along with their corresponding PDFs and statistics.

By invoking the principle of the minimum number of parameters for statistical description, as introduced by Fisher (1922), we show that the entire statistics of velocity and temperature increments can be analytically derived. This approach relies on a decomposition into bifurcation modes which, being non-observable quantities, are governed by quasi-probability distribution functions (quasi-PDFs). By admitting negative values, the latter provide the formal mathematical basis required to also represent local energy backscatter.

Notably, the analysis recovers the Kolmogorov law --specifically the scaling of the ratio between the velocity standard deviation and the Kolmogorov velocity as $R_\lambda^{1/2}$-- as a direct consequence of the non--observability of these modes. Furthermore, our analysis reveals that the bifurcation modes exhibit amplitudes whose third statistical moment scales as $R_\lambda^{-3}$. The results show excellent agreement with benchmark numerical and experimental data existing in the literature, confirming that the proposed bifurcation--based decomposition rigorously captures the multi--scale, intermittent nature of fully developed turbulence.
\end{abstract}

\begin{keyword}

Bifurcation Mode, Non--Observable Quasi--PDF, Kolmogorov Law, Intermittency
\end{keyword}

\end{frontmatter}

\newcommand{\no}{\noindent}
\newcommand{\be}{\begin{equation}}
\newcommand{\ee}{\end{equation}}
\newcommand{\bea}{\begin{eqnarray}}
\newcommand{\eea}{\end{eqnarray}}
\newcommand{\bc}{\begin{center}}
\newcommand{\ec}{\end{center}}

\newcommand{\calr}{{\cal R}}
\newcommand{\calv}{{\cal V}}

\newcommand{\bff}{\mbox{\boldmath $f$}}
\newcommand{\bfg}{\mbox{\boldmath $g$}}
\newcommand{\bfh}{\mbox{\boldmath $h$}}
\newcommand{\bfi}{\mbox{\boldmath $i$}}
\newcommand{\bfm}{\mbox{\boldmath $m$}}
\newcommand{\bfp}{\mbox{\boldmath $p$}}
\newcommand{\bfr}{\mbox{\boldmath $r$}}
\newcommand{\bfu}{\mbox{\boldmath $u$}}
\newcommand{\bfv}{\mbox{\boldmath $v$}}
\newcommand{\bfx}{\mbox{\boldmath $x$}}
\newcommand{\bfy}{\mbox{\boldmath $y$}}
\newcommand{\bfw}{\mbox{\boldmath $w$}}
\newcommand{\bfk}{\mbox{\boldmath $\kappa$}}

\newcommand{\bfA}{\mbox{\boldmath $A$}}
\newcommand{\bfD}{\mbox{\boldmath $D$}}
\newcommand{\bfI}{\mbox{\boldmath $I$}}
\newcommand{\bfL}{\mbox{\boldmath $L$}}
\newcommand{\bfM}{\mbox{\boldmath $M$}}
\newcommand{\bfS}{\mbox{\boldmath $S$}}
\newcommand{\bfT}{\mbox{\boldmath $T$}}
\newcommand{\bfU}{\mbox{\boldmath $U$}}
\newcommand{\bfX}{\mbox{\boldmath $X$}}
\newcommand{\bfY}{\mbox{\boldmath $Y$}}
\newcommand{\bfK}{\mbox{\boldmath $K$}}

\newcommand{\bfeta}{\mbox{\boldmath $\eta$}}
\newcommand{\bfrho}{\mbox{\boldmath $\rho$}}
\newcommand{\bfchi}{\mbox{\boldmath $\chi$}}
\newcommand{\bfphi}{\mbox{\boldmath $\phi$}}
\newcommand{\bfPhi}{\mbox{\boldmath $\Phi$}}
\newcommand{\bflambda}{\mbox{\boldmath $\lambda$}}
\newcommand{\bfxi}{\mbox{\boldmath $\xi$}}
\newcommand{\bfLambda}{\mbox{\boldmath $\Lambda$}}
\newcommand{\bfPsi}{\mbox{\boldmath $\Psi$}}
\newcommand{\bfomega}{\mbox{\boldmath $\omega$}}
\newcommand{\bfOmega}{\mbox{\boldmath $\Omega$}}
\newcommand{\bfeps}{\mbox{\boldmath $\varepsilon$}}
\newcommand{\bfepsn}{\mbox{\boldmath $\epsilon$}}
\newcommand{\bfzeta}{\mbox{\boldmath $\zeta$}}
\newcommand{\bfkappa}{\mbox{\boldmath $\kappa$}}
\newcommand{\bfsigma}{\mbox{\boldmath $\sigma$}}
\newcommand{\itPsi}{\mbox{\it $\Psi$}}
\newcommand{\itPhi}{\mbox{\it $\Phi$}}

\newcommand{\bint}{\mbox{ \int{a}{b}} }
\newcommand{\ds}{\displaystyle}
\newcommand{\Sum}{\Large \sum}



\bigskip

\section{Introduction \label{intro}}

The statistical description of fully developed turbulence remains one of the most formidable challenges in classical physics. Central to this endeavor is the concept of the energy cascade, a phenomenon originally envisioned by Richardson \cite{Richardson1922} and later formalized by Kolmogorov \cite{K41}, wherein kinetic energy is transferred from large-scale injections to small-scale dissipative structures. Within this framework, the small-scale statistics are assumed to become isotropic and universal at sufficiently high Reynolds numbers. A rigorous mathematical cornerstone of this theory is the Kolmogorov $4/5$ law, derived from the von K\'arm\'an--Howarth equation \cite{VonKarman1938}, which relates the third-order longitudinal structure function directly to the mean energy dissipation rate $\epsilon$.

As noted by Batchelor and Townsend \cite{Batchelor1949}, the non-zero value of the third-order moment implies a fundamental broken symmetry in the distribution of velocity increments, manifested as a persistent negative skewness. This skewness serves as the dynamical signature of the forward energy cascade and is intimately linked to the vortex stretching mechanism. However, decades of experimental and numerical research \cite{Sreenivasan1991, Anselmet1984, MoninYaglom1975} have demonstrated that the self-similar scaling predicted by K41 is violated by internal intermittency. Recent extreme-resolution Direct Numerical Simulations (DNS) \cite{Buaria2020, Iyer2020} have further elucidated how the dissipation field is concentrated on small-scale, high-intensity structures, reinforcing the refined hypotheses of Kolmogorov \cite{K62} and Oboukhov \cite{Oboukhov1962}.

The quest to quantify these departures led to the fractal and multifractal formalisms \cite{Mandelbrot1974, FrischParisi1985} and subsequent dynamical models \cite{Frisch1978, SheLeveque1994}. Contemporary investigations \cite{Meneveau2022, Johnson2023} have highlighted the persistent difficulty in reconciling the local geometry of vortex tubes with the global statistical scaling. For passive scalars, such as temperature, the Corrsin--Oboukhov theory \cite{Corrsin1951} often fails to capture the intense ramp-and-cliff structures and the extreme intermittency documented from classic studies \cite{Warhaft2000, ShraimanSiggia2000} to modern computational analyses \cite{Buaria2022}.

Although skewness and intermittency have occupied a central role in the scientific literature, to the best of the author’s knowledge, an integrated analysis of these phenomena through the lens of bifurcation mode non-observability and Fisher’s principle of minimum free parameters \cite{Fisher1922} has not yet received the attention it deserves. The present work addresses this gap by situating the study of the link between skewness and intermittency within the framework of the closure schemes for the von K\'arm\'an--Howarth and Corrsin equations previously developed by the author in \cite{deDivitiis2011, deDivitiis2014, deDivitiis2016, deDivitiis2026}.

Furthermore, a novel bifurcation analysis of the closed von K\'arm\'an--Howarth equation is proposed, which considers the route starting from fully developed turbulence toward non--chaotic regimes. This extends the discussion of previous works and represents an alternative point of view for studying the turbulent transition. According to this analysis, the closed von K\'arm\'an--Howarth equation is decomposed into several ordinary differential equations through the Taylor series expansion of the longitudinal velocity correlation. This procedure leads to an estimate of the Taylor-scale Reynolds number at the transition. This critical value is found to be approximately $10$, which is in good agreement with several experimental observations. Notably, this result aligns closely with the bifurcation analysis of the energy cascade presented in \cite{deDivitiis2011}, which provides a critical Reynolds number of $10.13$ assuming the route toward turbulence follows the Feigenbaum scenario \cite{Feigenbaum78, Eckmann81}.

Subsequently, by leveraging the Navier--Stokes and heat equations, we analytically define and estimate velocity and temperature fluctuations as variations occurring over a time interval during which the quadratic form associated with the local fluid deformation remains definite in sign.

Building upon these definitions and the aforementioned closed correlation equations, this work rigorously derives the structure function scaling and the statistical distributions that were previously introduced on a heuristic basis in \cite{deDivitiis2011, deDivitiis2014, deDivitiis2013}. This analysis, supported by a formal bifurcation review of the energy cascade \cite{deDivitiis2019}, suggests that while nonlinearity drives intermittency, it is the non-observability of bifurcation modes that constitutes the critical link for recovering Kolmogorov scaling. Importantly, the non-observability of these modes not only explains the Kolmogorov scaling and the intermittency increasing with the Taylor-scale Reynolds number, but above all, it allows for the determination of the internal structure functions for velocity and temperature differences, as well as the corresponding PDFs and statistics that closely match the existing experimental and numerical data in the literature.

By invoking Fisher's principle of parsimony \cite{Fisher1922}, we demonstrate that the complete statistics of velocity and temperature increments can be analytically derived through a decomposition into bifurcation modes. These modes are treated as non-observable quantities governed by quasi-PDFs. Notably, the proposed analysis reveals that the bifurcation modes exhibit a third-order statistical moment scaling as $R_\lambda^{-3}$. Consequently, the Kolmogorov scaling $\sqrt{R_\lambda}$ is recovered as a direct manifestation of this non-observability, showing excellent agreement with recent benchmark data \cite{Buaria2020, Yao2025}.

\bigskip

\section*{Theoretical Basis: Correlation Equations and Nondiffusive Closures}

The present analysis is strictly founded upon the closure schemes for the von K\'arm\'an--Howarth and Corrsin equations established by the author in \cite{deDivitiis2026}. In that work, the emergence of a Liouville spectral gap and the phenomenon of bifurcation-driven Lagrangian--Eulerian decoupling provided a physical-mathematical basis for nondiffusive turbulence closures. To ensure self-consistency, we recall the governing equations and the specific functional forms of the closures utilized herein.

In the context of homogeneous isotropic turbulence, the dynamical evolution of the longitudinal velocity correlation $f=\langle u_r u_r' \rangle_E/u^2$ and the temperature correlation $f_\theta = \langle \vartheta  \vartheta' \rangle_E/\theta^2$ are governed by:
\bea
\begin{array}{l@{\hspace{-0.cm}}l}
\ds \frac{\partial f}{\partial t} = 
\ds  \frac{K}{u^2} +
\ds 2 \nu  \left(  \frac{\partial^2 f} {\partial r^2} +
\ds \frac{4}{r} \frac{\partial f}{\partial r}  \right) +\frac{10 \nu}{\lambda_T^2} f, \\\\
\ds \frac{\partial f_\theta}{\partial t} = 
\ds  \frac{G}{\theta^2} +
\ds 2 \kappa  \left(  \frac{\partial^2 f_\theta} {\partial r^2} +
\ds \frac{2}{r} \frac{\partial f_\theta}{\partial r}  \right) +\frac{12 \kappa}{\lambda_\theta^2} f_\theta,
\end{array} 
\label{vk-h}
\eea
where  $u \equiv \sqrt{\langle u_r^2 \rangle_E}$, $\theta \equiv \sqrt{\langle \vartheta^2 \rangle_E}$,  $\left\langle .  \right\rangle_E$ denotes the average calculated over the Eulerian ensemble, and the quantities $\lambda_T \equiv \sqrt{-1/f''(0)}$ and $\lambda_\theta \equiv \sqrt{-2/f_\theta''(0)}$ are Taylor and Corrsin microscales, respectively.
The functions $K$ and $G$, which account for the turbulent energy cascade, are related to $k$ and $m^*$, namely the longitudinal triple velocity correlation and the mixed triple correlation between $u_r$ and $\vartheta$, according to 
\bea
\begin{array}{l@{\hspace{+0.0cm}}l}
\ds K(r) = u^3 \left( \frac{\partial }{\partial r} + \frac{4}{r} \right) 
k(r), 
\ \ \mbox{where} \ \ 
\ds k(r) = \frac{\langle u_r^2 u_r' \rangle_{E}}{u^3}, \\\\
\ds G(r) = 2 u \theta^2 \left( \frac{\partial }{\partial r} + \frac{2}{r} 
\right) m^*(r), 
\ \ \mbox{where} \ \ 
\ds m^*(r) = \frac{\langle u_r \vartheta \vartheta' \rangle_{E}}{\theta^2 u},
\end{array}
\eea
 Equations (\ref{vk-h}) are closed once both $K$ and $G$ are expressed solely in terms of $f$ and $f_\theta$.
Following the Liouville analysis of Ref. \cite{deDivitiis2026}, the closures for $K$ and $G$ are given by:
\bea
\begin{array}{l@{\hspace{+0.0cm}}l}
\ds K(r) =
 u^3 \sqrt{\frac{1-f}{2}} \ \frac{\partial f}{\partial r}, \\\\
\ds G(r) = 
 u \theta^2 \sqrt{\frac{1-f}{2}} \ \frac{\partial f_\theta}{\partial r},
\end{array}
\label{K}
\label{K closure}
\eea
These functional forms effectively replace the traditional eddy-diffusivity assumptions with a description based on the nonlinear interaction of bifurcation modes. As demonstrated in \cite{deDivitiis2026}, these closures are non-diffusive in nature, as they arise from the decoupling of Lagrangian trajectories in the presence of a spectral gap.

For the results obtained using Eqs. (\ref{K}), the reader is referred to the data reported in Refs. \cite{deDivitiis2011, deDivitiis2014, deDivitiis2012}. In particular, such Refs. demonstrate that Eqs. (\ref{K}) provide an accurate description of the energy cascade, reproducing negative values of the skewness of velocity and temperature increments 
\bea
\begin{array}{l@{\hspace{+0.0cm}}l}
\ds H_{u 3}(r) \equiv 
\frac{\langle (\Delta u_r)^3 \rangle }{\langle (\Delta u_r)^2 \rangle^{3/2}} 
=
\frac{6 k(r)}{(2(1-f(r)))^{3/2}} \\\\
\ds H_{\theta 3}(r) \equiv 
\frac{\langle (\Delta \vartheta)^2 \Delta u_r \rangle }
{\langle (\Delta \vartheta)^2 \rangle {\langle (\Delta u_r)^2 \rangle}^{1/2}}=
\frac{4 m^*}{ 2(1-f_\theta(r)) (2(1-f(r)))^{1/2}} 
\end{array}
\label{H3(r)}
\eea
 and, in particular,
\bea
\begin{array}{l@{\hspace{+0.0cm}}l}
\ds H_{u 3}(0) = \lim_{r \rightarrow 0} H_{u 3}(r) = - \frac{3}{7}, \\\\
\ds H_{\theta 3}(0) = \lim_{r \rightarrow 0} H_{\theta 3}(r) = - \frac{1}{5}, 
\end{array}
\label{H3(0)}
\eea
 in agreement with the literature \cite{Chen92, Orszag72, Panda89, Anderson99, Carati95, Kang2003}, with the Kolmogorov law, and with temperature spectra consistent with the theoretical arguments of Kolmogorov, Obukhov--Corrsin, and Batchelor \cite{Batchelor_2, Batchelor_3, Obukhov}, as well as with experimental \cite{Gibson, Mydlarski} and numerical \cite{Rogallo, Donzis} results.

This work takes these closed equations (\ref{vk-h}), (\ref{K}) and Eq. (\ref{H3(0)}) as a point of departure to investigate the interplay between skewness and intermittency. By applying Fisher’s principle of parsimony \cite{Fisher1922} to the statistics generated by these closures, we demonstrate that the non-observability of the underlying bifurcation modes is the fundamental mechanism responsible for the recovery of the Kolmogorov scaling and the emergence of non-Gaussian statistics.

Although the second of Eqs. (\ref{K}) is formulated for temperature correlation equation, the present analysis applies, without loss of generality, to any passive scalar with diffusive transport.

\bigskip

\section*{Mathematical Framework: Quasi-PDFs, Non-observable Bifurcation Modes and Fisher Parsimony Principle}

The analytical derivation of the statistics for velocity and temperature increments---the primary physical observables of the system---rests upon a decomposition into discrete bifurcation modes. Within this framework, while the state variables, velocity $\mathbf{u}$ and temperature $\vartheta$, are directly accessible to measurement, the individual bifurcation modes governing their multi-scale evolution remain inherently latent. 

A fundamental distinction arises from their governing dynamics: while the observable field $\mathbf{u}$ satisfies the Navier--Stokes equations, the individual bifurcation modes do not. These modes represent mathematical segregations of the fluid state that possess physical reality only in their collective combination. Consequently, such modes are fundamentally \textit{non-observable}. According to Liouville's theorem, the observable field $\mathbf{u}$ obeys a classical, positive-definite distribution; conversely, the non-observable modes, emerging from this formal decomposition, are governed by quasi-probability density functions (quasi-PDFs) which may exhibit negative values \cite{Feynman87, Burgin2009, Burgin2010}.

{
The \textbf{bifurcation modes} of the Navier-Stokes equations are formally defined as \textbf{finite-amplitude Lyapunov modes} (nonlinear modes). These are evaluated along the reference trajectory at its intersections with the \textbf{hypersurface where the Jacobian determinant vanishes}, or in the immediate vicinity of these numerous phase-space points. 
Unlike infinitesimal perturbations in the tangent space, these modes $\mathbf{v}$ evolve according to the \textbf{nonlinear Lyapunov equation}:
\begin{equation}
    \dot{\mathbf{v}} = \mathbf{N}(\mathbf{u} + \mathbf{v}) - \mathbf{N}(\mathbf{u})
\end{equation}
where $\mathbf{u}$ is the reference velocity field and $\mathbf{N}$ is the Navier-Stokes operator. The \textbf{finite amplitude} is determined by the global norm of $\mathbf{v}$, which is physically scaled to the \textbf{minimum distance between sequential bifurcation points} along the trajectory in phase space. 

By bypassing the linear approximation of the tangent space, this framework captures the essential \textbf{nonlinear characteristics} of finite perturbations. Since these nonlinear modes are orthonormal or orthonormalizable through the Gram-Schmidt procedure, the velocity field can be locally represented \textbf{exactly} as a linear combination of such modes. Furthermore, the field can be effectively approximated by projecting it onto the subspace spanned by modes associated with \textbf{non-negative Lyapunov exponents} ($\lambda_k \ge 0$):
\begin{equation}
    \mathbf{u}(t, \mathbf{x}) \approx \sum_{k: \lambda_k \ge 0} a_k(t) \mathbf{v}_k(\mathbf{x})
\end{equation}
 Near a bifurcation, the \textbf{Lyapunov basis}---defined by the orthonormalization of the modes---exhibits \textbf{discontinuous jumps in its orientation} within the phase space. These orientation singularities represent the topological signature of the bifurcation, acting as the primary driver for intermittency.

The \textbf{non-observability} of these modes is a structural consequence of their nature:
\begin{enumerate}
    \item \textit{Nonlinear Finite Nature:} These modes describe the internal transition mechanism between metastable states. They represent the "pathway" of the bifurcation rather than a standalone observable state.
    \item \textit{Basis Indivisibility:} Because the physical measurement (e.g., a hot-wire probe or PIV) only detects the aggregate field $\mathbf{u}$, the individual modes $\mathbf{v}_k$ remain \textbf{structurally inseparable} from the observable flow. They function as a hidden decomposition basis, necessitating the use of \textbf{quasi-PDFs} to account for the intermittent fluctuations they induce at small scales.
\end{enumerate}}

The global statistical behavior emerges from the collective contribution of these latent modes, governed by the closure of the von K\'arm\'an--Howarth and Corrsin equations \cite{deDivitiis2011, deDivitiis2014, deDivitiis2026}.

The mathematical necessity for quasi-PDFs to admit negative excursions stems from the physics of non-observability. This has profound implications for the energy cascade: specifically, the coexistence of negative and positive regions within the quasi-PDF provides the formal basis for representing local \textit{backscatter}---the upscale transfer of kinetic energy. This phenomenon is a hallmark of the non-linear dynamics of the Navier--Stokes equations in fully developed turbulence. Accordingly, the total observed skewness of the velocity increments, $H_{u 3}(r)$, can be analytically mapped onto the underlying quasi-PDF structure.

\subsection*{Non-observability and the Fisher Parsimony Principle}

The present theory posits that bifurcation modes are \textit{non-observable} entities whose existence is inferred solely through the macroscopic statistics of resolved increments. Adhering to the principle of parsimony \cite{Fisher1922}, which mandates a sufficient statistical description via a minimal set of parameters, we demonstrate that turbulent complexity is minimized through the selection of a specific, irreducible set of modes.

This non-observability constitutes the critical conceptual link between intrinsic non-linearity and the recovery of Kolmogorov scaling. By requiring the statistical description to be compatible with a minimal set of free parameters in the Fisherian sense, the model constrains the PDFs of $\Delta u_r$ and $\Delta \vartheta$. This constraint, in conjunction with the Navier--Stokes non-linearity, leads directly to the analytical prediction of intermittency, which increases monotonically with $R_\lambda$. Furthermore, it recovers the Kolmogorov law, whereby the standard deviation of observable increments scales as $\langle (\Delta u_r)^2 \rangle^{1/2} / u_K \propto \sqrt{R_\lambda}$ ($u_K$ being the Kolmogorov velocity). 

Crucially, while the Navier--Stokes non-linearity inherently implies the existence of intermittency, it is the non-observability of the bifurcation modes that enforces the Kolmogorov scaling. In this perspective, intermittency that increases with $R_\lambda$ is not an \textit{ad hoc} correction, but rather the unique statistically consistent state emerging from the non-observability of bifurcation modes and the constant skewness of $\Delta u_r$, under the rigorous constraint of parsimony. Moreover, this mechanism also induces an intermittency of the temperature difference which rises accordingly with the P\'eclet number.

\bigskip

\section{Bifurcation analysis and Transition Reynolds number: from fully developed turbulence toward non--chaotic regimes.
}

The transition toward fully developed turbulence traditionally proceeds from non--chaotic regimes via intermediate stages \cite{Ruelle71, Feigenbaum78, Pomeau80, Eckmann81}, corresponding to successive bifurcations at comparable Reynolds scales. In a departure from this classical route to chaos, this section investigates the inverse route: we examine the loss of stability as the system evolves from fully developed homogeneous isotropic turbulence toward a non--chaotic state. Along this path, characterized by a progressive reduction in the Taylor--scale Reynolds number, we analyze the bifurcations inherent in the closed von K\'arm\'an--Howarth equation. This framework allows for the estimation of a critical threshold, $R_{T}^*$, which represents the minimum $R_\lambda$ necessary to sustain a fully developed, homogeneous, and isotropic turbulent state. The bifurcation analysis of the velocity correlation is formulated by expanding $f(t, r)=\Sum_k f^{(k)}_0 r^k/k!$ into an even Taylor series. The evolution of the resulting coefficients is governed by a system of coupled equations, where each coefficient corresponds to a specific dynamical degree of freedom
\bea
\left\lbrace 
\begin{array}{l@{\hspace{+0.0cm}}l}
\ds  \frac{du}{dt} = -5 \nu \frac{u}{\lambda_T^2}, \\\\
\ds  \frac{d \lambda_T}{dt} = -\frac{u}{2} + \frac{\nu}{\lambda_T}
\left( \frac{7}{3} f^{IV}_0 \lambda_T^4-5\right), \\\\
\ds \frac{d f^{IV}_0}{dt} = ..., \\
\ds ...  \\
\ds \frac{d f^{(n)}_0}{dt} = ..., \\
\ds ... 
\end{array}\right.
\label{dec vkh}
\eea
Such equations can be written by introducing the infinite dimensional state vector 
\bea
\ds {\bf f} \equiv \left( u, \lambda_T, f_0^{IV},....f_0^{(n)},...\right).
\eea
which represents the state of the longitudinal velocity correlation. 
Therefore, Eqs. (\ref{dec vkh}), formally written as
\bea
\ds \dot{\bf f} ={\bf F}({\bf f}, \nu)
\label{kh bif}  
\eea
are equivalent to the closed von K\'arm\'an--Howarth equation.
Equation (\ref{kh bif}) defines a bifurcation problem where $\nu$ plays the role of control parameter. Thus, this bifurcation analysis studies the variations of $\bf f$ caused by $\nu$ according to  
\bea
\ds {\bf F}({\bf f}, \nu)={\bf F}({\bf f}_0, \nu_0)
\eea
starting from ${\bf f}_0$, $\nu_0$.
For $\nu > \nu_0$, $\bf f$ is formally calculated through the implicit functions inversion theorem 
\bea
\ds {\bf f} = {\bf G}({\bf f}_0, \nu_0, \nu) \equiv {\bf f}_0 - \int_{\nu_0}^{\nu} 
\left( \nabla_{\bf f} {\bf F} \right)^{-1} \frac{\partial {\bf F}}{\partial \nu} \ d \nu
\label{bif Y}
\eea
where $\nabla_{\bf f} {\bf F}$ is the jacobian $\partial {\bf F}/\partial {\bf f}$. 
A bifurcation of Eq. (\ref{kh bif}) happens when this jacobian is singular, i.e.
\bea
\ds \det \left( \nabla_{\bf f} {\bf F} \right) =0
\label{bif F}
\eea
In the limit of vanishingly small \(\nu_{0}\) (corresponding to sufficiently large \(R_{T}\)), the energy cascade dominates viscous effects, ensuring the non-singularity of the Jacobian \(\nabla _{\mathbf{f}}\mathbf{F}\). As \(\nu \) increases, the state vector \(\mathbf{f}\) evolves smoothly according to Eq. (\ref{bif Y}) until dissipation progressively overwhelms the energy cascade, leading to the primary bifurcation defined by condition (\ref{bif F}). At this critical juncture, a 'hard' loss of stability occurs, marking the collapse of fully developed turbulence toward non-chaotic regimes \cite{Arnold92}. Consequently, the threshold \(R_{T}^{*}\) is identified as the value at bifurcation that maximizes the largest real part of the eigenvalues of \(\nabla _{\mathbf{f}}\mathbf{F}\) \cite{Arnold13, Arnold92}, consistent with the prescribed average kinetic energy \(u^{2}\), namely
\bea
\begin{array}{l@{\hspace{+0.0cm}}l}
\ds R_\lambda^*  \  \ \vert  \ \ \sup_k\left\lbrace\Re(l_k) \right\rbrace=\mbox{max}, \\\\
\ds \det \left( \nabla_{\bf f} {\bf F} \right) =0, \\\\
\ds  u^2=\mbox{given}
\end{array}
\label{opt jac}
\eea
this is a constrained extremum condition, where $l_k$, $k$=1, 2,... are the eigenvalues of $\nabla_{\bf f}{\bf F}$.
We assume that the phase space of Eq. (\ref{kh bif}), while inherently infinite-dimensional, can be effectively projected onto a finite-dimensional manifold. Within this framework, Eqs. (\ref{kh bif}) are mapped into the dynamical systems paradigm pioneered by Ruelle and Takens \cite{Ruelle71}, allowing the governing equations to be analyzed through the lens of classical bifurcation theory for ordinary differential equations. Mathematically, this transition is justified by the convergence of the functional series representing the solution; provided that the truncation satisfies uniform convergence criteria (e.g., the Weierstrass M-test), the bifurcation analysis remains asymptotically consistent with the full operator dynamics. The omitted degrees of freedom, residing in the series "tail," do not qualitatively alter the stability or the branching evolution of the solutions.

From the perspective of the laminar--turbulent transition, this reduction is further substantiated by the existence of a local center manifold. The onset of instability is driven by a discrete set of critical modes that dictate the qualitative changes in flow topology, while the remaining infinite degrees of freedom exhibit sufficiently strong decay to be treated as enslaved to the primary dynamics. This effectively compresses the infinite-dimensional phase space into its most influential components without compromising the underlying bifurcation structure.

Physically, the energy cascade primarily drives the temporal reduction of $\lambda_T$ while $f$ maintains a self-similar profile. This suggests that, in the proximity of the transition, the variables $u$ and $\lambda_T$ provide a sufficient descriptor for the evolution of $f$. Consequently, the state vector ${\bf f}$ is truncated to its first two components, corresponding to the $r^0$ and $r^2$ terms in Eqs. (\ref{dec vkh}). Leveraging this self-similarity, the infinite--dimensional space is replaced by a low--dimensional manifold, where the state vector reduces to:
\bea
\ds {\bf f} \equiv \left( u, \lambda_T\right),
\eea
Thus, the jacobian $\nabla_{\bf f} {\bf F}$ is reduced to
\bea
\begin{array}{l@{\hspace{+0.0cm}}l}
\nabla_{\bf f}{\bf F}
=
\left(\begin{array}{cc}
\ds  \frac{\partial \dot{u}}{\partial u} & \ds \frac{\partial \dot{u}}{\partial \lambda_T} \\\\
\ds  \frac{\partial \dot{\lambda_T}}{\partial u} & \ds \frac{\partial \dot{\lambda_T}}{\partial \lambda_T} 
\end{array}\right)
\end{array}
\eea
whose determinant is
\bea
\ds \det\left( \nabla_{\bf f}{\bf F}\right) = -\frac{5 \nu^2}{\lambda_T^2} 
\left(7 f_0^{IV} \lambda_T^2  + \frac{5}{\lambda_T^2}\right) +
5 \nu \frac{u}{\lambda_T^3} 
\label{det grafF}
\eea
where $f_0^{IV}$ plays the role of a variable in the constrained extremum condition (\ref{opt jac}).
Equation (\ref{det grafF}) implies that $\det(\nabla_{\bf f}{\bf F}) > 0$ for sufficiently small $\nu > 0$. The emergence of a bifurcation requires the vanishing of the Jacobian determinant. As $\nu$ increases, the ratio $\det(\nabla_{\bf f}{\bf F})/\nu$ monotonically decreases until it reaches a singular point. To determine the critical threshold $R_\lambda^*$, we eliminate $f_0^{IV}$ by invoking the bifurcation condition $\det(\nabla_{\bf f}{\bf F}) = 0$ in Eq. (\ref{det grafF}), yielding:
\bea
\ds f_0^{IV} = \frac{1}{7 \lambda_T^2 \nu} \left( \frac{u}{\lambda_T}-5\frac{\nu}{\lambda_T^2}\right) 
\eea
The resulting singular Jacobian is given by:
\bea
\begin{array}{l@{\hspace{+0.0cm}}l}
\nabla_{\bf f}{\bf F}
=
\left(\begin{array}{cc}
\ds  -5  \nu/\lambda_T^2 & \ds 10 \nu u/\lambda_T^3  \\\\
\ds  -1/2 & \ds u/\lambda_T
\end{array}\right)
\end{array}
\eea
which admits the eigenvalues $l_1, l_2$ and corresponding eigenvectors ${\bf y}_1, {\bf y}_2$:
\bea
\begin{array}{l@{\hspace{+0.0cm}}l}
\ds l_1 = 0, \quad {\bf y}_1 = \left( u, \frac{\lambda_T}{2} \right)  \\\\
\ds l_2 = \frac{u^2}{\nu}\left( \frac{1}{R_\lambda} - \frac{5}{R_\lambda^2} \right) , \quad {\bf y}_2 = \left(u, R_\lambda \frac{\lambda_T}{10} \right)
\end{array}
\eea
The real eigenvalue $l_2$ remains positive for $R_\lambda > 5$ and attains its maximum, 
$l_{2, {max}} = 5 \nu/\lambda_T^2$, at $R_\lambda = 10$. Consequently, we estimate the critical Taylor-scale Reynolds number as:
\bea
\begin{array}{l@{\hspace{+0.0cm}}l}
R_\lambda^* = 10
\end{array}
\label{RT min}
\eea

A further characteristic threshold is identified when both eigenvalues vanish at $R_\lambda = 5$, marking the onset of the decaying turbulence regime. In this limit, the evolution of $f$ and $\lambda_T$ is well-approximated by:
\bea
\begin{array}{l@{\hspace{+0.0cm}}l}
\ds \frac{d \lambda_T}{dt} \simeq 0, \  \
\ds f \simeq \exp \left( -\frac{1}{2} \left( \frac{r}{\lambda_T}\right)^2 \right)  
\end{array}
\eea
leading to $f^{IV}_0 \lambda_T^4 \simeq 3$ and $R_\lambda \simeq 4$, which is consistent with the aforementioned estimation.

The value $R_\lambda^*$ represents the lower bound for sustaining fully developed, homogeneous, and isotropic turbulence, thereby establishing the order of magnitude for the transition. While the actual route to chaos involves intermediate Navier--Stokes bifurcations where isotropy and homogeneity may not yet be established, this analysis suggests that during transition, $R_\lambda$ spans the range:
\bea
\begin{array}{l@{\hspace{+0.0cm}}l}
\ds 4 \lesssim R_\lambda \lesssim 10
\end{array}
\eea
The estimated value $R_\lambda^* = 10$ is in excellent agreement with bifurcation analyses of the turbulent energy cascade \cite{deDivitiis_3}, which show that if the cascade follows the Feigenbaum scenario \cite{Feigenbaum78, Eckmann81}, a critical Taylor--scale Reynolds number of approximately $10.13$ emerges after three successive bifurcations.

In summary, the estimation of $R_\lambda^*$ is predicated on the local self--similarity induced by the closures (\ref{K}), which justifies the reduction of the infinite-dimensional system to the primary two-equation manifold in (\ref{dec vkh}).

\bigskip

\section{Velocity and temperature fluctuations}

This section aims to derive formal expressions for Eulerian velocity and temperature fluctuations in fully developed turbulence, providing a rigorous basis for estimating the statistical properties of these fields. The derivation leverages the Lyapunov theory of Lagrangian trajectories, as established in \cite{deDivitiis2026}, by employing a dual framework of referential and spatial coordinates.

For analytical convenience, the Navier--Stokes and heat equations are cast into the following divergence form:
\bea
\left\lbrace 
\begin{array}{l}
\ds \frac{\partial {\bf u}}{\partial t} = \frac{1}{\rho} \nabla_{\bf x} \cdot \hat{\bf T} \\\\
\ds \frac{\partial \vartheta }{\partial t} = - \frac{1}{c \rho}\nabla_{\bf x} \cdot \hat{\bf q}  
\end{array} 
\right.
\label{NST_eq_div}
\eea
where the total flux tensors, $\hat{\bf T}$ and $\hat{\bf q}$, are defined as:
\bea
\begin{array}{l@{\hspace{+0.5cm}}l}
\ds \hat{\bf T} = {\bf T} - \rho {\bf u} \otimes {\bf u}, & \ds \hat{\bf q} = {\bf q} +  c \rho \ {\bf u} \ \vartheta
\end{array}
\eea
In this context, ${\bf T}$ and ${\bf q}$ denote the stress tensor and heat flux, respectively, which satisfy the Navier--Fourier constitutive relations:
\bea
\begin{array}{l@{\hspace{+0.5cm}}l@{\hspace{+0.5cm}}l}
\ds {\bf T} = -{\bf I} p + {\bf T}_v, & \ds {\bf q} = - \kappa \nabla_{\bf x} \vartheta, & \ds {\bf T}_v = \mu \left( \nabla_{\bf x} {\bf u} + \nabla_{\bf x} {\bf u}^T \right)
\label{Tq}
\end{array}
\eea
where the spatial gradients are evaluated with respect to the Eulerian coordinates $\bf x$ \cite{Truesdell77}. Here, $\bf I$ is the identity tensor, ${\bf T}_v$ is the viscous stress tensor, while $\mu, \kappa, c$, and $\rho$ represent the dynamic viscosity, thermal conductivity, specific heat, and density, respectively---all treated as constant quantities. 
The pressure $p$ is determined via the continuity equation and expressed as a functional of the velocity field:
\bea
\ds  p(t, {\bf x}) = \frac{\rho}{4 \pi} \int \frac{\partial^2 u_i' u_j'}{\partial x_i' \partial x_j'} \ \frac{d V({\bf x}')}{\vert {\bf x}' - {\bf x} \vert} 
\label{pressure}
\eea
This formulation explicitly accounts for non-local effects \cite{Tsinober2009}, effectively reducing the Navier--Stokes equations to an integro--differential system. 

Velocity and temperature fluctuations are defined as variations occurring over a time interval $(t_0, t)$ such that the quadratic form $Q$ associated with the local deformation gradient $\partial {\bf x}/\partial {\bf X}$ maintains a constant sign, where $\bf X$ denotes the referential (Lagrangian) coordinate. 
{ The sign-definiteness of the quadratic form associated with the deformation gradient offers a direct link to the \textbf{Q-criterion} used in coherent structure identification. When the form is \textbf{sign-definite}, the local flow is dominated by the strain-rate tensor $S_{ij}$, characterizing regions of high scalar dissipation. Conversely, an \textbf{indefinite form} indicates the prevalence of the rotation tensor $\Omega_{ij}$, identifying the cores of turbulent vortices. Our approach thus relates the emergence of intermittency to the topological transitions between these two states, which are triggered by the orientation jumps of the Lyapunov basis. In terms of Reynolds decomposition $\mathbf{u} = \langle \mathbf{u} \rangle + \mathbf{u}'$, our approach identifies fluctuations not merely as deviations from the mean, but as topological transitions where the bifurcation modes switch the signature of $Q$. This provides a dynamical basis for the "internal structure" of intermittency that standard Reynolds decomposition obscures.}

By applying the Lyapunov theory of Lagrangian trajectories, Eqs. (\ref{NST_eq_div}) are transformed using the referential coordinates and the local fluid deformation:
\bea 
\begin{array}{l@{\hspace{+0.8cm}}l}
\ds \frac{\partial {\bf u}}{\partial t} = \frac{1}{\rho}\frac{\partial \hat{\bf T}}{\partial {\bf X}} \left( \frac{\partial {\bf x}}{\partial {\bf X}}\right)^{-1}, & \ds \frac{\partial \vartheta}{\partial t} = - \frac{1}{c \rho}\frac{\partial \hat{\bf q}}{\partial {\bf X}}  \left( \frac{\partial {\bf x}}{\partial {\bf X}}\right)^{-1} 
\end{array}
\label{rate1}
\eea
The inverse deformation gradient
\bea
 \ds {\bf G} =  \left( \frac{\partial {\bf x}}{\partial {\bf X}}\right)^{-1}
\eea
is defined such that ${\bf G}(t_0) = {\bf I}$. 

The simultaneous adoption of referential and spatial coordinates enables the factorization of $\partial {\bf u}/\partial t$ and $\partial \vartheta/\partial t$ into a product of two matrices characterized by disparate time scales: one governed by the Eulerian velocity and temperature fields, and the other representing the local fluid distortion. Following \cite{deDivitiis2026}, $\bf G$ evolves according to Lyapunov theory, while the Lagrangian gradients $\partial \hat{\bf T}/\partial {\bf X}$ and $\partial \hat{\bf q}/\partial {\bf X}$ exhibit a slow-growth property. Consequently, fluctuations are defined over a time interval during which the quadratic form associated with $\bf G$ remains definite in sign.
Hence, the velocity and temperature fluctuations are expressed as:
\bea
\begin{array}{l@{\hspace{+0.8cm}}l}
\ds {\bf u} = \frac{1}{\rho} \int_{t_0}^t \left( \frac{\partial \hat{\bf T}}{\partial {\bf X}}\right)_\tau
\left( \frac{\partial {\bf x}}{\partial {\bf X}}\right)_\tau^{-1} d \tau, & 
\ds \vartheta = - \frac{1}{c \rho} \int_{t_0}^t \left( \frac{\partial \hat{\bf q}}{\partial {\bf X}}\right)_\tau   
\left( \frac{\partial {\bf x}} {\partial {\bf X}}\right)_\tau^{-1} d \tau
\end{array}
\label{fluct int}
\eea
where the subscript $\tau$ indicates the time at which the quantities are evaluated, and ${\bf X}$ denotes the Lagrangian coordinate calculated at time $t_0$ according to ${\bf X}={\bfchi}^{-1}(\tau, {\bf x})$ with $\tau \in (t_0, t)$ \cite{Truesdell77}.
As the quadratic form associated with $\partial {\bf x}/\partial {\bf X}$ preserves its sign within the interval $(t_0, t)$, the integral mean value theorem can be applied to Eqs. (\ref{fluct int}), yielding:
\bea
\begin{array}{l@{\hspace{+0.8cm}}l} 
\ds \exists \ t_U, t_\Theta \in (t_0, t) \ \vert \   {\bf u} = \frac{1}{\rho} \left( \frac{\partial \hat{\bf T}}{\partial {\bf X}} \right)_{t_U} {\bf W}, & \ds \vartheta = -\frac{1}{c \rho}\left( \frac{\partial \hat{\bf q}}{\partial {\bf X}} \right)_{t_\Theta} {\bf W}, \ \ 
\end{array}
\label{fluct u t}
\eea
where the matrix ${\bf W}$ is defined as:
\bea
\ds {\bf W} = \int_{t_0}^t {\bf G} \ d \tau 
\eea
In conclusion, for developed turbulence, Eqs. (\ref{fluct u t}) represent exact relations derived by integrating spatial and referential coordinates with local deformation. This provides a robust analytical framework for describing fluctuation dynamics and constitutes a fundamental starting point for characterizing the statistics of velocity and temperature fields in developed turbulence.

\bigskip

\section{Velocity and temperature difference and non-observable bifurcation modes}

In developed turbulence, the longitudinal velocity and temperature increments, $\Delta u_r = ({\bf u}(t, {\bf x}') - {\bf u}(t, {\bf x})) \cdot {\bf r}/r$ and $\Delta \vartheta = \vartheta(t, {\bf x}') - \vartheta(t, {\bf x})$ with ${\bf r} = {\bf x}' - {\bf x}$, are of paramount importance, as they encapsulate the dynamics of the energy cascade, intermittency, and turbulent dissipation. This section employs the Navier--Stokes and heat equations, as summarized by (\ref{fluct u t}), to analyze the analytical structure of $\Delta u_r$ and $\Delta \vartheta$ in fully developed homogeneous isotropic turbulence. This is achieved by leveraging the previously established Lagrangian Lyapunov framework alongside a specific statistical decomposition that accounts for bifurcation-induced effects.
Specifically, within the present framework, $\Delta u_r$ and $\Delta \vartheta$ are decomposed into bifurcation modes. These modes represent non-observable quantities, each statistically characterized by a quasi-PDF. Such distributions may exhibit impulsive profiles and admit negative values, reflecting their nature as signed measures rather than classical probability densities. We demonstrate that a specific choice of these quasi-PDFs identically satisfies Eqs. (\ref{H3(r)})--(\ref{H3(0)}), which define the dimensionless third-order statistical moments of $\Delta u_r$ and $\Delta \vartheta$.

To analyze this, we first examine the impact of Navier--Stokes bifurcations on $\Delta u_r$ and $\Delta \vartheta$.
 Using the fluctuation relations derived in Eq. (\ref{fluct u t}), the increments are expressed in terms of the instantaneous velocity and temperature fields as:
\bea
\begin{array}{l}
\ds \Delta u_r = \frac{1}{\rho}\left( \frac{\partial \hat{T}_{i j}}{\partial x_{0 k}} \right)_{t_U'}' W_{j k}' - \frac{1}{\rho} \left( \frac{\partial \hat{T}_{i j}}{\partial x_{0 k}} \right)_{t_U} W_{j k} \\\\
\ds \Delta \vartheta = - \frac{1}{c \rho} \left( \frac{\partial \hat{q}_j}{\partial x_{0 k}} \right)_{t_\Theta'}' W_{j k}' +  \frac{1}{c \rho}\left( \frac{\partial \hat{q}_j}{\partial x_{0 k}} \right)_{t_\Theta} W_{j k}
\end{array}
\label{fluct du dt}
\eea
The sequence of bifurcations encountered during the fluid motion results in a continuous doubling of the velocity field into multiple components, denoted as ${\hat{\bf v}}_k$ for $k=1, 2, \dots$. Each bifurcation introduces new functions whose characteristics are statistically independent of the velocity field at preceding times. Consequently, the velocity field is decomposed as:
\bea
\ds {\bf u}(t, {\bf x}) = \sum_k {\hat{\bf v}}_k(t, {\bf x})
\eea
It is crucial to remark that while the total field ${\bf u}(t, {\bf x})$ satisfies the Navier--Stokes equations, the individual components $\hat{\bf v}_k$ do not. These functions represent bifurcation modes, mathematical segregations of the fluid state that physically exist only in combination. As such, these modes are fundamentally non--observable. According to Liouville's theorem, the observable field ${\bf u}$ follows a classical, positive-definite distribution. Conversely, the non-observable modes $\hat{\bf v}_k$, resulting from this mathematical segregation, are governed by quasi-PDFs which may exhibit negative values \cite{Feynman87, Burgin2009, Burgin2010}.

The form of these quasi-PDFs is dictated by the Navier--Stokes bifurcations and the isotropy hypothesis. Regarding the former, for pressure and inertia to drive significant variations in velocity autocorrelation, each mode $\hat{\bf v}_k = (\hat{v}_1, \hat{v}_2, \hat{v}_3)$ must follow a highly non-symmetric distribution, characterized by:
\bea
\ds \frac{\vert \langle \hat{v}_{k i}^3 \rangle \vert}{\langle \hat{v}_{k i}^2 \rangle^{3/2}} \gg 1, \quad i = 1, 2, 3
\label{c_xi_1}
\eea
Simultaneously, under the isotropy hypothesis, ${\bf u}$ would typically follow a Gaussian distribution \cite{Batchelor53}, wherein inertia and pressure do not contribute to the time derivative of the third statistical moment. To reconcile this with the energy cascade, the third moments of $\hat{\bf v}_k$ are expected to dominate the higher-order statistics:
\bea
\ds \frac{\vert \langle \hat{v}_{k i}^3 \rangle \vert}{\langle \hat{v}_{k i}^2 \rangle^{3/2}} \gg \frac{\vert \langle \hat{v}_{k i}^4 \rangle \vert}{\langle \hat{v}_{k i}^2 \rangle^2}, \quad i = 1, 2, 3 
\label{c_xi_2}
\eea
This framework justifies the representation of the velocity field ${\bf u}$ and the passive scalar $\vartheta$ as linear combinations of stochastic variables $\xi_k$, which effectively encapsulate the doubling effect characteristic of bifurcations:
\bea
\begin{array}{l}
\ds {\bf u} = \sum_k {\bf U}_k \xi_k, \\\\
\ds \vartheta = \sum_k \Theta_k \xi_k
\end{array}
\label{stat decomp}
\eea
where the spatial modes ${\bf U}_k$ and $\Theta_k$ satisfy the incompressibility constraint $\nabla_{\bf x} \cdot {\bf U}_k = 0$, and $\xi_k$ denote independent centered stochastic variables. The quasi-PDF ${\hat \delta}(\xi_k)$, consistent with Eqs. (\ref{c_xi_1}) and (\ref{c_xi_2}), is formulated via a singular Gram--Charlier expansion truncated at the third order, developed around a nascent delta function $\delta_e(\xi_k)$:
\bea
\ds {\hat \delta}(\xi_k) = \delta_e(\xi_k) - \frac{\left\langle \xi_k^3 \right\rangle }{3!} \delta^{(3)}(\xi_k)
\label{Gram Charlier}
\eea
In this construction, the standard deviation of $\delta_e(\xi_k)$ is assumed to be infinitesimal yet non-vanishing, thereby ensuring the formal applicability of the Central Limit Theorem (CLT) under summation. This specific expansion characterizes a physical regime in which the mode exhibits a dominant energy flux ($\left| \left\langle \xi_k^3 \right\rangle \right| \ggg 1$ and $\left| \left\langle \xi_k^3 \right\rangle \right| \ggg \left\langle \xi_k^2 \right\rangle$) while maintaining a vanishing local energy contribution, as evidenced by the infinitesimal variance. The inclusion of the third-order derivative of the Dirac delta implies that the mode serves exclusively as a vehicle for the turbulent cascade, redistributing energy across the spectrum without contributing to the local energy density-a defining feature of non-observable bifurcation modes in fully developed turbulence.
\bea
\begin{array}{l@{\hspace{-0.cm}}l}
\left\langle \xi_k \right\rangle = 0, \  \ \ 
\left\langle \xi_i \xi_j \xi_k \right\rangle= \left\lbrace 
\begin{array}{l@{\hspace{-0.cm}}l}
\ds q \ne 0, \ \forall \ i=j=k  \\\\
\ds 0 \ \ \mbox{else} 
\end{array}\right. 
\end{array}
\label{prop1}
\eea
where $q$, providing the skewness of $\xi_k$ k=1, 2..., satisfies to
\bea
\begin{array}{l@{\hspace{-0.cm}}l}
\vert q \vert \ggg \left\langle \xi_k^n \right\rangle, \forall n \ne 3, \ \  k=1, 2, ...
\end{array}, 
\label{prop2}
\eea
{{\bf Remark}. \it While the formal representation (\ref{stat decomp}) might superficially resemble a {\it Karhunen Lo\'eve Expansion} (KLE) \cite{loeve1977}, it departs from it in two fundamental aspects that are crucial for the description of the Navier-Stokes bifurcation cascade.
First, unlike the observable modes in a standard KLE, the variables $\xi_k$ here represent \textit{non-observable} bifurcation states. According to the \textit{Center Manifold Theory} \cite{carr1981}, near a bifurcation point, the system's dynamics are slaved to critical modes. In our framework, these modes are segregated: they represent pure energy transfer states rather than directly measurable velocity fluctuations.
Second, the stochastic nature of each mode is described by a \textit{quasi-PDF} rather than a classical probability density. Following the intuition of Feynman \cite{Feynman87} on negative probabilities for intermediate states, each $\xi_k$ is governed by a singular Gram-Charlier expansion (\ref{Gram Charlier}).
This formulation allows for the mathematical representation of intermittency and local energy flux without violating the global Gaussianity of the system, which is recovered in the limit of large $N$ through the \textit{Lindeberg-Feller} condition \cite{Ventsel}.}

Through the decomposition in Eq. (\ref{stat decomp}), we demonstrate that the negative value of $H^{(3)}_u(r)$ carries profound implications for the statistical properties of $\Delta u_r$ and $\Delta \vartheta$, particularly concerning the emergence of intermittency as the Taylor-scale Reynolds and P\'eclet numbers increase. To investigate this, we first derive the analytical fluctuations of $u_i$ and $\vartheta$ in terms of the bifurcation variables $\xi_k$ by substituting Eq. (\ref{stat decomp}) into Eq. (\ref{fluct u t}). Hence, dimensionless eulerian velocity and temperature fluctuations read as 
\bea
\left\lbrace 
\begin{array}{l}
\ds u_i = \sum_j \sum_k A^{(i)}_{j k} \xi_j \xi_k + \frac{1}{R_\lambda} \sum_k a^{(i)}_k \xi_k, \quad i=1, 2, 3 \\\\
\ds \vartheta = \sum_j \sum_k B_{j k} \xi_j \xi_k + \frac{1}{Pe} \sum_k b_k \xi_k
\end{array}
\right.
\label{fluct u t xi}
\eea 
where the equations were non--dimensionalized using the standard deviations of velocity and temperature, along with the corresponding Taylor and Corrsin correlation scales, thus $R_\lambda= u \lambda_T/\nu$,  $Pe = R_\lambda Pr$,  $Pr = \nu/\alpha$, $\alpha =\kappa/( c \rho) $.
In this formulation, the quadratic forms $\sum \sum A^{(i)}_{j k} \xi_j \xi_k$ and the linear terms $1/R_\lambda \sum a^{(i)}_k \xi_k$ represent the contributions of inertial/pressure forces and fluid viscosity, respectively. Similarly, for the passive scalar $\vartheta$, the terms arise from convective transport and thermal conduction.

Under the hypothesis of turbulent isotropy, the velocity $u_i$ and temperature $\vartheta$ are expected to behave as Gaussian stochastic variables \cite{Batchelor53, Ventsel, Lehmann99}. Physically, this implies that the vast number of independent bifurcation modes $\xi_k$ contributing to the global field satisfy the Lindeberg condition, a necessary and sufficient requirement for the Central Limit Theorem (CLT) to hold \cite{Ventsel, Lehmann99}. Crucially, this condition remains valid even for singular quasi-PDFs, as the infinitesimal contribution of each individual non-observable mode ensures a convergent Gaussian macroscopic state.
Specifically, the second terms on the right-hand side (RHS) of Eqs. (\ref{fluct u t xi}) are sums of uncorrelated variables described by quasi-PDFs, where both $a^{(i)}_k$ and $b_k$ satisfy the Lindeberg condition. The first terms are quadratic forms where, due to isotropy, the structure of the matrices $A^{(i)}_{j k}$ and $B_{j k}$ is such that their spectral radii are negligible with respect to the corresponding Frobenius norms. This constitutes a necessary and sufficient condition for $u$ and $\vartheta$ to be distributed according to Gaussian PDFs \cite{deJong1987}.

However, the CLT does not extend to the increments $\Delta u_i$ and $\Delta \vartheta$. Since these quantities result from the difference between two highly correlated Gaussian fields, the subtraction process effectively filters out the uncorrelated Gaussian background, thereby exposing the underlying singular structure of the bifurcation modes. In this scale-selective regime, the Lindeberg condition is no longer satisfied, and the PDF of the increments deviates significantly from a Gaussian distribution, manifesting the observed intermittency. To characterize this statistical behavior, the fluctuations of the increments are expressed as:
\bea
\begin{array}{l}
\ds \Delta u_r({\bf r}) = \sum_j \sum_k \Delta A_{j k} \xi_j \xi_k + \frac{1}{R_\lambda} \sum_k \Delta a_k \xi_k, \\\\
\ds \Delta \vartheta({\bf r}) = \sum_j \sum_k \Delta B_{j k} \xi_j \xi_k + \frac{1}{Pe} \sum_k \Delta b_k \xi_k
\end{array}
\label{fluct du dt xi}
\eea
Conversely, the second and third statistical moments of $\Delta u_r$ and $\Delta \vartheta$ are coupled via the energy cascade mechanism, in accordance with Eqs. (\ref{H3(0)}) and (\ref{H3(r)}). 
By substituting the decompositions (\ref{fluct du dt xi}) into Eqs. (\ref{H3(0)}) and taking into account that $\vert \left\langle \xi_k^3 \right\rangle\vert \ggg \left\langle \xi_k^n \right\rangle, \forall n \ne 3$, it is found that Eqs. (\ref{H3(0)}) are identically satisfied by the quasi-PDF:
\bea
\ds {\hat \delta}(\xi_k) = \delta_e(\xi_k) - \frac{\left\langle \xi_k^3 \right\rangle }{3!} \delta^{(3)}(\xi_k)
\label{Gram Charlier 2}
\eea
if and only if the following condition holds:
\bea
\ds \left\langle \xi_k^3 \right\rangle = \frac{C}{R_\lambda^3}
\label{xi3}
\eea
where $C$ is an arbitrary constant, and provided that:
\bea
\ds \frac{\ds \sum_k \Delta A_{k k} \Delta B^2_{k k}}{\ds \sum_k \Delta a_{k k} \Delta b^2_{k k}} \sim \left( \frac{1}{Pr}\right)^2
\eea
Within the theoretical framework of the Fisher parsimony principle, these results provide 
robust confirmation that Eqs. (\ref{Gram Charlier 2}) and (\ref{xi3}) constitute the 
necessary and sufficient conditions for the velocity and temperature gradients, 
$\partial u_r/\partial r$ and $\partial \vartheta/\partial r$ as defined through
Eqs.(\ref{fluct du dt xi}), to exhibit constant, non-zero negative skewness values 
in strict accordance with Eqs. (\ref{H3(0)}). Notably, these equations represent 
quasi--PDFs of non-observable quantities whose analytical structure accounts for 
the energy flux of the cascade while rigorously preserving the mean kinetic energy 
of the system.

\bigskip

\section{Statistics of velocity and temperature difference}

To evaluate the statistics of $\Delta u_r$ and $\Delta \vartheta$, 
 the matrices $\Delta A_{j k}$ and $\Delta B_{j k}$ in Eq. (\ref{fluct du dt xi}), are first decomposed into their  respective symmetric parts ($S_{u j k}, S_{\theta j k}$), associated with the local strain and dissipation, and antisymmetric parts ($\Omega_{u j k}, \Omega_{\theta j k}$), associated with the local vorticity.
\bea
\begin{array}{l}
\ds \Delta A_{j k} = \sum_{i=1}^3 \left( A^{(i)}_{j k}({\bf x} + {\bf r}) - A^{(i)}_{j k}({\bf x}) \right) \frac{r_i}{r} \equiv S_{u j k} + \Omega_{u j k} \\\\
\ds \Delta a_k = \sum_{i=1}^3 \left( a^{(i)}_k({\bf x} + {\bf r}) - a^{(i)}_k({\bf x}) \right) \frac{r_i}{r} \\\\
\ds \Delta B_{j k} = B_{j k}({\bf x} + {\bf r}) - B_{j k}({\bf x}) \equiv S_{\theta j k} + \Omega_{\theta j k} \\\\
\ds \Delta b_k = b_k({\bf x} + {\bf r}) - b_k({\bf x})
\end{array}
\label{dec0}
\eea
 This decomposition allows us to trace the emergence of non--Gaussian tails directly to the non-linear interaction of bifurcation modes at the scales of the energy cascade.
The terms involving $\Omega_{u j k}$ and $\Omega_{\theta j k}$ yield a null contribution to Eqs. (\ref{fluct du dt xi}). Consequently, $S_{u j k}$ and $S_{\theta j k}$ are decomposed into their spectral diagonal components and a residual coupling matrix by applying a spectral shift to the main diagonal:
\bea
\begin{array}{l@{\hspace{-0.cm}}l}
\ds S_{X j k} = \Lambda^+_{X j  k} - \Lambda^-_{X j k} + R_{X j k}, \quad X = u, \theta
\end{array}
\eea
where $\Lambda^+_{X j  k}$ and $-\Lambda^-_{X j k}$ denote diagonal matrices whose non-zero entries correspond to the positive and non-positive eigenvalues of $S_{X j k}$, respectively, while $R_{X j k}$ represents the residual matrices. The latter satisfy the Gerschgorin circle theorem, according to which the absolute value of the diagonal residual $\left| R_{X i i} \right|$ is bounded by the sum of the absolute values of the off-diagonal elements in the corresponding row of ${\bf R}_X$. 

Accordingly, the velocity and temperature increments can be expressed as:
\bea
\begin{array}{l}
\ds \Delta u_r({\bf r}) = \sum_{j k} \left( \Lambda^+_{u j  k}-\Lambda^-_{u j  k} \right)  \xi_j \xi_k + \sum_{j k} R_{u j k} \xi_j \xi_k + \frac{1}{R_\lambda} \sum_k \Delta a_k \xi_k, \\\\
\ds \Delta \vartheta({\bf r}) = \sum_{j k} \left( \Lambda^+_{\theta j  k}-\Lambda^-_{\theta j  k} \right)\xi_j \xi_k + \sum_{j k} R_{\theta j k} \xi_j \xi_k + \frac{1}{Pe} \sum_k \Delta b_k \xi_k
\end{array}
\label{fluc dec}
\eea
Consequently, the centered fluctuations of $\Delta u_r$ and $\Delta \vartheta$ are reformulated as functions of the variables $\eta_{X}$, $\eta_{X +}$, and $\eta_{X -}$ (for $X = u, \theta$):
\bea
\begin{array}{l@{\hspace{-0.cm}}l}
\ds \Delta u_r = \eta_u + 
                 \left( \eta^2_{u+} -  \left\langle \eta^2_{u+} \right\rangle \right) - 
                 \left( \eta^2_{u-} -  \left\langle \eta^2_{u-} \right\rangle \right), \\\\
\ds \Delta \vartheta = \eta_\theta + 
           \left( \eta^2_{\theta+} -\left\langle \eta^2_{\theta+} \right\rangle \right) - 
            \left( \eta^2_{\theta-} -\left\langle \eta^2_{\theta-} \right\rangle \right),
\end{array}
\eea
where these variables are defined as follows:
\bea
\begin{array}{l@{\hspace{-0.cm}}l}
\ds \eta_{X +} = \Sum_k \left( \Lambda^+_{X k k}\right)^{1/2} \xi_k,\ \ \eta_{X -} = \Sum_k \left( \Lambda^-_{X k k}\right)^{1/2} \xi_k, \ \ X= u, \theta,
\end{array}
\label{gauss var}
\eea
\bea
\begin{array}{l@{\hspace{-0.cm}}l}
\ds \eta_u = \frac{1}{R_\lambda}\Sum_i \Delta a_i \xi_i + \Sum_{j k} F_{u j k} \xi_j \xi_k, \\\\
\ds \eta_\theta = \frac{1}{Pe} \Sum_i \Delta b_i \xi_i + \Sum_{j k} F_{\theta j k} \xi_j \xi_k, 
\end{array}
\label{gauss var 2}
\eea
with
\bea
\begin{array}{l@{\hspace{-0.cm}}l}
\ds F_{X j k} = R_{X j k} + 
                \left( \Lambda^-_{X j j} \Lambda^-_{X k k}\right)^{1/2}   -
                \left( \Lambda^+_{X j j} \Lambda^+_{X k k}\right)^{1/2}, \ \ X = u, \theta
\end{array}
\label{matrix F}
\eea

We now demonstrate that $\eta_X$, $\eta_{X -}$, and $\eta_{X +}$ asymptotically approach uncorrelated Gaussian variables. Indeed, as per Eq. (\ref{gauss var}), $\eta_{X +}$ and $\eta_{X -}$ are sums of terms derived from two distinct sets of uncorrelated stochastic variables. Thus, $\eta_{X +}$ and $\eta_{X -}$ emerge as centered uncorrelated stochastic variables, such that $\langle \eta_{X +} \rangle = \langle \eta_{X -} \rangle = 0$. Furthermore, since the $\xi_k$ are statistically independent, even if they are described by quasi-PDFs, the application of the Central Limit Theorem (CLT) to Eq. (\ref{gauss var}) ensures that both $\eta_{X +}$ and $\eta_{X -}$ converge to centered Gaussian random variables.

Regarding $\eta_X$, the following considerations apply: the first terms on the right-hand side (RHS) of Eqs. (\ref{gauss var 2}) are sums of uncorrelated variables described by quasi-PDFs, where both $\Delta a_i$ and $\Delta b_i$ satisfy the Lindeberg condition. The second terms are quadratic forms where, due to the structure of the matrices ${\bf F}_X$ in Eqs. (\ref{matrix F}) and the fact that ${\bf R}_X$ satisfies the Gerschgorin theorem, the spectral radii of ${\bf F}_X$ are negligible compared to their corresponding Frobenius norms. This constitutes a necessary and sufficient condition for $\eta_X$ to be distributed according to a Gaussian PDF \cite{deJong1987}.

Hence, the statistics of $\Delta u_r$ and $\Delta \vartheta$ can be reduced to structure functions of the independent centered Gaussian stochastic variables $\zeta_X$, $\zeta_{X+}$, and $\zeta_{X-}$, obtained by normalizing $\eta_X$, $\eta_{X+}$, and $\eta_{X-}$, respectively, such that $\langle \zeta_X^2 \rangle = \langle \zeta_{X+}^2 \rangle = \langle\zeta_{X-}^2\rangle = 1$:
\bea
\begin{array}{l@{\hspace{-0.cm}}l}
\Delta u_r = L_u \zeta_u + S^+_u (\zeta_{u+}^2-1) -S^-_u(\zeta_{u-}^2-1), \\\\
\Delta \vartheta = L_\theta \zeta_\theta + S^+_\theta (\zeta_{\theta+}^2-1) -S^-_\theta(\zeta_{\theta-}^2-1),
\end{array}
\eea
where $L_X$, $S_X^+$, and $S_X^-$ are normalization coefficients introduced to ensure that $\zeta_X$, $\zeta_{X+}$, and $\zeta_{X-}$ possess unit standard deviation. Thus:
\bea 
\begin{array}{l@{\hspace{-0.cm}}l}
\ds L_u \zeta_u = \Sum_{i j} F_{u i j}  \xi_i \xi_j + \frac{1}{R_\lambda}\Sum_k \Delta a_k \xi_k, \\\\ 
\ds L_\theta \zeta_\theta = \Sum_{i j} F_{\theta i j}  \xi_i \xi_j + \frac{1}{Pe}\Sum_k \Delta b_k \xi_k,
\end{array}
\label{Lx}
\eea
where the $r$-dependent parameters $L_X$, $S^-_X$, and $S^+_X$ remain to be determined. 

In this regard, it is noteworthy that, according to Fisher \cite{Fisher1922}, in a regime of fully developed isotropic turbulence within an infinite domain, the number of parameters required to describe the statistics of $\Delta u_r$ and $\Delta \vartheta$ should be the minimum compatible with the prescribed physical quantities defining the state of motion (e.g., average kinetic energy, temperature standard deviation, and correlation functions). Furthermore, the evolution equation for the velocity pair correlation $f$ \cite{VonKarman1938} requires knowledge of the third-order correlations $k$. Consequently, within the framework of estimating the evolution of $f$ in fully developed homogeneous isotropic turbulence, the knowledge of $f$ and $k$ alone is considered necessary and sufficient to determine the statistics of $\Delta u_r$. This implies that $S_u^+$ is proportional to $S_u^-$ through a quantity independent of $r$, namely:
\bea
\begin{array}{l@{\hspace{-0.cm}}l}
\ds S^+_u(r) = \chi S^-_u(r) \equiv \chi S_u(r)
\end{array}
\eea
where $\chi < 1$ is a function of $R_\lambda$ representing the skewness of $\Delta u_r$. Accordingly, $S_u$ and $L_u$ are determined as functions of $f$ and $k$ once $\chi = \chi(Re)$ is identified. Regarding the temperature difference, turbulence isotropy dictates that the skewness of $\Delta \vartheta$ must vanish, yielding:
\bea
\ds S^+_\theta(r) = S^-_\theta(r) \equiv S_\theta(r)
\eea
The structure functions for $\Delta u_r$ and $\Delta \vartheta$ thus read:
\bea
\begin{array}{l@{\hspace{-0.cm}}l}
\Delta u_r = L_u \zeta_u + S_u \left( \chi \left( \zeta_{u+}^2-1\right)  -\left( \zeta_{u-}^2-1\right) \right) , \\\\
\Delta \vartheta= L_\theta \zeta_\theta + S_\theta \left( \zeta_{\theta+}^2 - \zeta_{\theta-}^2\right),
\end{array}
\label{du dt}
\eea
Finally, invoking the principle of the minimum number of parameters, the ratio $\Psi_\theta(r) \equiv S_\theta/L_\theta$ is assumed to be proportional to $\Psi_u(r) \equiv S_u/L_u$ via a coefficient solely dependent on the Prandtl number:
\bea
\begin{array}{l@{\hspace{-0.cm}}l}
\ds \Psi_\theta(r) = \sigma(Pr) \Psi_u(r)
\end{array}
\eea
where $\sigma(Pr)$ is a function to be determined.

At this stage of the analysis, we demonstrate that in the regime of fully developed turbulence, $L_u$ and $L_\theta$ are functions of $R_\lambda$ and $Pe$, respectively, scaling as $L_u \propto R_\lambda^{-1/2}$ and $L_\theta \propto Pe^{-1/2}$. Indeed, from Eq. (\ref{Lx}), we obtain:
\bea
\begin{array}{l@{\hspace{-0.cm}}l}
\ds L_u^2 = 
\Sum_{i j k l}  F_{u i j}  F_{u k l} \left\langle \xi_i \xi_j \xi_k \xi_l \right\rangle + 
\frac{2}{R_\lambda} \Sum_k F_{u k k} \Delta a_{u k} \left\langle \xi_k^3 \right\rangle +
\frac{1}{R_\lambda^2} \Sum_k \Delta a^2_k \left\langle \xi_k^2\right\rangle, \\\\
\ds L_\theta^2 = 
\Sum_{i j k l}  F_{\theta i j}  F_{\theta k l} \left\langle \xi_i \xi_j \xi_k \xi_l \right\rangle + 
\frac{2}{Pe} \Sum_k F_{\theta k k} \Delta b_k \left\langle \xi_k^3 \right\rangle +
\frac{1}{Pe^2} \Sum_k \Delta b^2_k \left\langle \xi_k^2\right\rangle,
\end{array}
\label{Lx2}
\eea
Since $\xi_k$ represent bifurcation modes described by Gram--Charlier quasi-PDFs, the condition $\vert \left\langle \xi_k^3 \right\rangle \vert \gg 1, \left\langle \xi_i \xi_j \xi_k \xi_l \right\rangle$ holds. Consequently, the first and third terms on the right-hand side of Eq. (\ref{Lx2}) become negligible, implying that $L_u$ and $L_\theta$ admit the following functional forms:
\bea
\begin{array}{l@{\hspace{-0.cm}}l}
\ds L_u = \frac{M_u(r)}{\sqrt{R_\lambda}}, \\\\
\ds L_\theta = \frac{M_\theta(r)}{\sqrt{Pe}}.
\end{array}
\eea
where $M_u(r)$ and $M_\theta(r)$ are $r$-dependent functions that do not exhibit direct dependence on $R_\lambda$ and $Pe$. 

Hence, the dimensionless increments $\Delta u_r$ and $\Delta \vartheta$, normalized by their respective standard deviations, are expressed as functions of $R_\lambda$ and $Pe$:
\bea
\begin{array}{l@{\hspace{-0.cm}}l}
\ds \frac{\Delta u_r}{\sqrt{ \langle (\Delta u_r)^2 \rangle}} = 
\frac{\zeta_u + \Psi_u (\chi(\zeta_{u+}^2-1)-(\zeta_{u-}^2-1))}{\sqrt{1+2 \Psi_u^2(1+ \chi^2) }}, 
\\\\ 
\ds  \Psi_u (r) = \frac{S_u(r)}{L_u(r)}=\Phi(r) \sqrt{R_\lambda}, \\\\
\ds \frac{\Delta \vartheta}{\sqrt{ \langle (\Delta \vartheta)^2 \rangle}} = 
\frac{\zeta_\theta + \Psi_\theta (\zeta_{\theta+}^2-\zeta_{\theta-}^2)}{\sqrt{1+4 \Psi_\theta^2 }}, \\\\
\ds \Psi_\theta(r) = \frac{S_\theta(r)}{L_\theta(r)}=\Phi(r) \sqrt{Pe} 
\end{array}
\label{struct funs}
\eea
which identifies the relationship $\sigma = \sqrt{Pr}$. Eqs. (\ref{struct funs}) define the specific structure functions governing the statistics of $\Delta u_r$ and $\Delta \vartheta$.

It is noteworthy that the term involving $\Psi_u$ accounts for the intermittency of $\Delta u_r$. Furthermore, according to Eqs. (\ref{fluct u t}) and (\ref{fluct du dt}), we have $\Psi_u(0) \sim (u/u_k)^2 (\ell_k/\lambda_T)$, where $u$ and $u_k$ denote the velocity standard deviation and the Kolmogorov velocity, respectively, while $\ell_k$ and $\lambda_T$ represent the Kolmogorov scale and the Taylor microscale. 

Remarkably, the present approach, by leveraging non-observable bifurcation modes and their corresponding quasi-PDFs, recovers the Kolmogorov scaling law, which states that $u/u_k \sim (u/u_k)^2 (\ell_k/\lambda_T) \sim \sqrt{R_\lambda}$. This suggests that the non-observability of bifurcation modes constitutes the fundamental conceptual link between the intermittency of $\Delta u_r$ and the Kolmogorov law. Without invoking such non-observability, while intermittency could still be identified, the derivation of the Kolmogorov law would remain inaccessible.

Now, assuming $\chi=\chi(R_\lambda)$ is known, $L_u$ and $S_u$ can be expressed as functions of $\langle \Delta u_r^2 \rangle$ and $\langle \Delta u_r^3 \rangle$, where the latter is determined by adopting the proposed closure (\ref{K}). In fact, $L_u$ and $S_u$ are related to these moments through Eq. (\ref{du dt}):
\bea
\begin{array}{l@{\hspace{-0.cm}}l}
\ds \left\langle (\Delta u_r)^3 \right\rangle = 6 u^3 k = 8 S_u^3 (\chi^3-1), \\\\
\ds \left\langle (\Delta u_r)^2 \right\rangle = 2 u^2 (1-f) = L_u^2+ 2 S_u^2 (\chi^2+1),
\end{array}
\eea
Consequently, $L_u$, $S_u$, and $\Phi$ are expressed in terms of $f(r)$ and $k(r)$ as follows:
\bea
\begin{array}{l@{\hspace{-0.cm}}l}
\ds S_u(r) =  \left(  \frac{3/4}{\chi^3-1}\right)^{1/3} u \ k(r)^{1/3}, \\\\
\ds L_u(r)=\sqrt{2} \ u  \sqrt{1-f(r)-(1+\chi^2) \left( \frac{3/4}{\chi^3-1}\right)^{2/3} \ k(r)^{2/3}}, \\\\
\ds \Phi = \frac{S_u}{L_u} \frac{1}{\sqrt{R_\lambda}}
\end{array}
\label{S_u L_u}
\eea
In the expression for $L_u(r)$ in Eqs. (\ref{S_u L_u}), the argument of the square root must be strictly positive, leading to the following implicit condition for $\chi$:
\bea
\begin{array}{l@{\hspace{-0.cm}}l}
\ds \frac{1+ \chi^2}{\left( \chi^3-1\right)^{2/3} } \le \frac{1}{2} \left( \frac{56}{3} \right)^{2/3} 
\end{array}
\label{chi}
\eea
incorporating the proposed closure (\ref{K}). Solving inequality (\ref{chi}) for $\chi$ yields the upper bound:
\bea
\begin{array}{l@{\hspace{-0.cm}}l}
\ds \chi \le \chi_\infty = 0.8659...
\end{array}
\eea

Regarding the temperature difference, we have:
\bea
\ds \frac{\left\langle \left( \Delta u_r\right)^2 \right\rangle} {\left\langle \left( \Delta \vartheta \right)^2 \right\rangle} \equiv 
\frac{u^2}{\theta^2} \ \frac{1-f}{1-f_\theta} =
\frac{L_u^2}{L_\theta^2} \ \frac{1+2\Psi_u^2(1+\chi^2)}{1+4\Psi_\theta^2}
\label{phi_theta}
\eea
Thus, Eq. (\ref{phi_theta}) allows for the calculation of $L_\theta$ in terms of the other quantities:
\bea
\begin{array}{l@{\hspace{-0.cm}}l}
\ds L_\theta= L_u \frac{\theta}{u} \sqrt{\frac{1-f_\theta}{1-f}}
\sqrt{\frac{1+2 \Phi^2 R_\lambda (1+\chi^2)}{1+4 \Phi^2 Pe}}
\end{array}
\label{phi_theta b}
\eea
In Eqs. (\ref{phi_theta b}) and (\ref{S_u L_u}), the function $\chi = \chi(R_\lambda)$ must be identified; furthermore, $\Phi(r)$ depends on the specific profile of $f(r)$. Due to the constancy of $H^{(3)} _u(0)$ and according to the Fisher principle, $\Phi(0)$ is assumed to be a constant independent of $R_\lambda$.

The PDFs of $\Delta u_r$ and $\Delta \vartheta$ are formally derived via the Frobenius--Perron equation \cite{Nicolis95}, considering that $\zeta_X$, $\zeta_{X-}$, and $\zeta_{X+}$ are independent, identically distributed centered Gaussian variables such that $\langle \zeta_X^2 \rangle = \langle \zeta_{X-}^2 \rangle = \langle \zeta_{X+}^2 \rangle = 1$ for $X = u, \theta$:
\bea
\begin{array}{l@{\hspace{-0.cm}}l}
\ds F_{u}(\Delta u_r') = \int_\zeta \int_{\zeta_{-}} \int_{\zeta_{+}} P(\zeta, \zeta_{-}, \zeta_{+}) \
\delta (\Delta u_r'-\Delta u_r(\zeta, \zeta_{-}, \zeta_{+})) \ \ d\zeta \ d\zeta_{-} \ d\zeta_{+}, \\\\
\ds F_{\theta}(\Delta \vartheta') = \int_\zeta \int_{\zeta_{-}} \int_{\zeta_{+}} P(\zeta, \zeta_{-}, \zeta_{+}) \
\delta (\Delta \vartheta'-\Delta \vartheta(\zeta, \zeta_{-}, \zeta_{+})) \ d\zeta \ d\zeta_{-} \ d\zeta_{+},
\end{array}
\label{Frobenius-Perron}
\eea
where $\delta$ is the Dirac delta function and $P(\zeta, \zeta_{-}, \zeta_{+})$ is the trivariate Gaussian PDF:
\bea
\ds P(\zeta, \zeta_{-}, \zeta_{+}) = \frac{1}{\sqrt{(2 \pi)^3}} \exp\left(-\frac{\zeta^2+\zeta_{-}^2+\zeta_{+}^2}{2}\right),
\label{gaussian}
\eea
with $\Delta u_r(\zeta, \zeta_{-}, \zeta_{+})$ and $\Delta \vartheta(\zeta, \zeta_{-}, \zeta_{+})$ determined by Eqs. (\ref{struct funs}).

\begin{figure}[h!]
\centering
\includegraphics[scale=.4, angle=0]{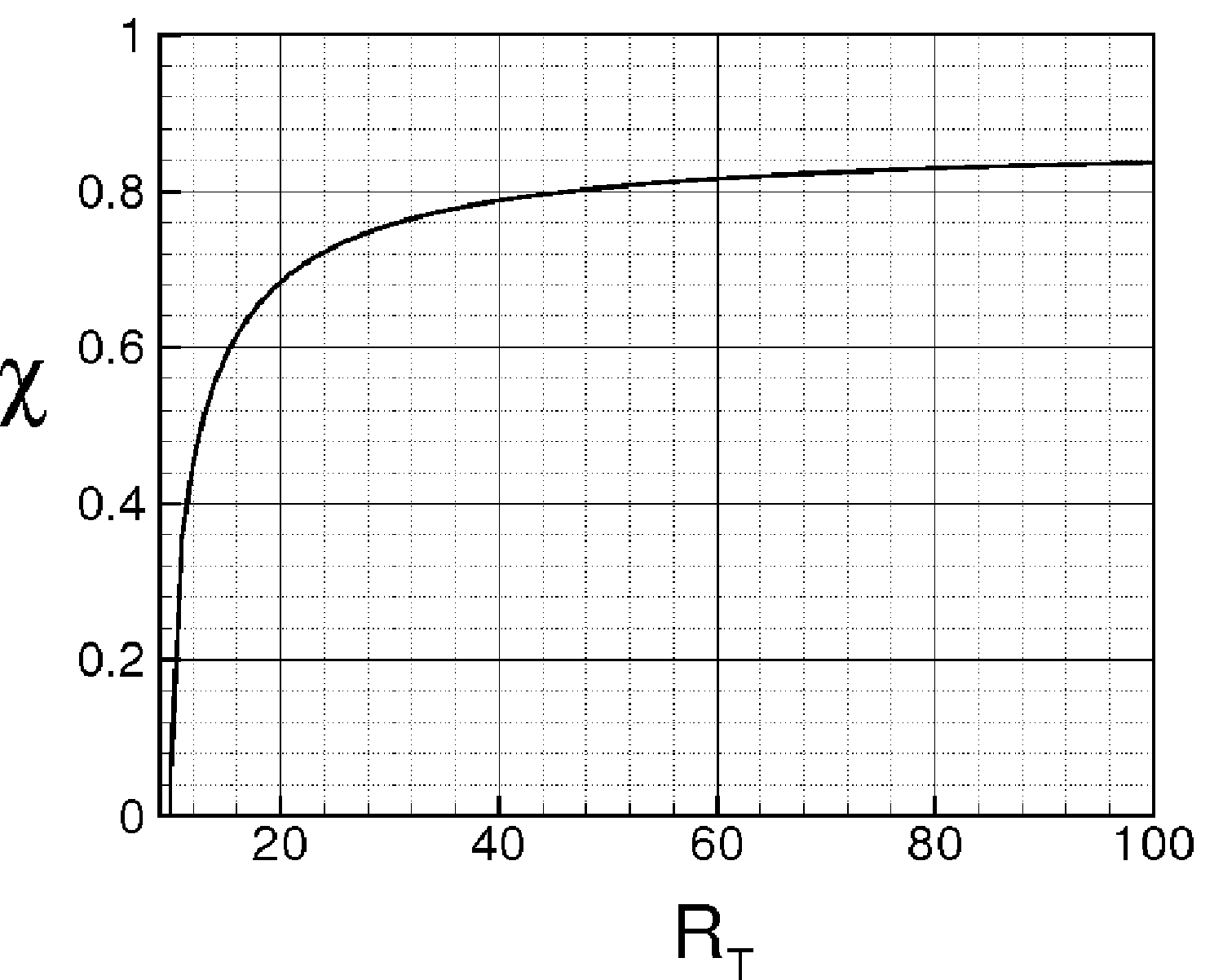}
\caption{Characteristic Function $\chi$=$\chi(R_\lambda)$}
\label{fig_chi_re}
\end{figure}

In essence, the statistics of $\Delta u_r$ and $\Delta \vartheta$ can be inferred from the proposed statistical decomposition (\ref{stat decomp}), which accounts for bifurcation effects in isotropic turbulence. This results in non-Gaussian statistics where the absolute values of the dimensionless moments increase with $R_\lambda$ and $Pe$. Specifically, the dimensionless statistical moments are calculated as:
\bea
\begin{array}{l@{\hspace{+0.2cm}}l}
\ds H_u^{(n)} \equiv \frac{\left\langle (\Delta u_r )^n \right\rangle}
{\left\langle (\Delta u_r)^2 \right\rangle^{n/2} }
= \\\\
\ds \frac{1} {(1+2(1+\chi^2)  \Psi_u^2)^{n/2}} 
\ds \sum_{k=0}^n 
\left(\begin{array}{c}
n  \\
k
\end{array}\right)  \Psi_u^k
 \langle \zeta_u^{n-k} \rangle 
  \langle (\chi (\zeta_{u+}^2-1)  - (\zeta_{u-}^2-1) )^k \rangle, \\\\
\ds H_\theta^{(n)} \equiv \frac{\left\langle (\Delta \vartheta )^n \right\rangle}
{\left\langle (\Delta \vartheta)^2 \right\rangle^{n/2} }
= 
\ds \frac{1} {(1+4  \Psi_\theta^2)^{n/2}} 
\ds \sum_{k=0}^n 
\left(\begin{array}{c}
n  \\
k
\end{array}\right)  \Psi_\theta^k
 \langle \zeta_\theta^{n-k} \rangle 
  \langle (\zeta_{\theta+}^2 - \zeta_{\theta-}^2 )^k \rangle, 
\end{array}
\label{Tm1} 
\eea
where $\Phi(0)$ and $\chi=$ $\chi(R_\lambda)$ remain to be identified. To this end, we first analyze the statistics of $\partial u_r / \partial r$, which, according to the proposed Lyapunov analysis, exhibits a constant skewness $H_u^{(3)}(0) = -3/7$. From Eqs. (\ref{Tm1}), we obtain:
\bea
\ds H_u^{(3)}(r) = \frac{8 \Psi_u^3(\chi^3-1)}{(1+ 2 \Psi_u^2 (1+\chi^2))^{3/2}} 
\eea
and taking the limit $r \rightarrow 0$:
\bea
\ds H_u^{(3)}(0) = \frac{8 \Psi_u^3(0)(\chi^3-1)}{(1+ 2 \Psi_u^2(0) (1+\chi^2))^{3/2}} 
\label{H30}
\eea
Accordingly, $\chi = \chi(R_\lambda)$ is implicitly defined as a function of $\Phi(0) \sqrt{R_\lambda}$. Eq. (\ref{H30}) shows that $\chi(R_\lambda)$ is a monotonically increasing function of $R_\lambda$, which, for $H_u^{(3)}(0) = -3/7$, reaches the limit $\chi_\infty = 0.8659...$ as $R_\lambda \rightarrow \infty$. For sufficiently small $R_\lambda$, $\chi(R_\lambda)$ may become negative. However, in fully developed turbulence, the PDF of $\partial u_r / \partial r$ must exhibit non-Gaussian tails as $\partial u_r / \partial r \rightarrow \pm \infty$, requiring $\chi$ to be positive. Thus, the limit $\chi = 0$ is assumed to occur at $R_\lambda = R_\lambda^* = 10$, representing the threshold for homogeneous isotropic turbulence. This allows for the identification of $\Phi(0)$ via Eq. (\ref{H30}):
\bea
\ds \Phi(0) = \frac{1}{\sqrt{R_\lambda^*}} \sqrt{ \frac{{H_{u 0}^{(3)}}^{2/3}}{4-2{H_{u 0}^{(3)}}^{2/3}}} = 0.1409...
\label{phi0}
\eea 
The resulting implicit variation law $\chi = \chi(R_\lambda)$ is illustrated in Fig. \ref{fig_chi_re}.
To validate the value of $\Phi(0)$ calculated in Eq. (\ref{phi0}), we consider that the order of magnitude scales as $\Phi(0) \sim (u/u_K)^2 (\ell_K/\lambda_T)$. Given that:
\bea
\begin{array}{l@{\hspace{-0.cm}}l}
\ds \frac{\lambda_T}{\ell_K} = 15^{1/4} \sqrt{R_\lambda}, \quad \frac{u_K \ell_K}{\nu} = 1, \quad R_\lambda = \frac{u \lambda_T}{\nu}  
\end{array}
\label{lambda_ell}
\eea
it follows that:
\bea
\ds \Phi(0) \sim \left( \frac{u}{u_K}\right)^2 \frac{\ell_K}{\lambda_T} = \frac{1}{\sqrt{15 \sqrt{15}}} \simeq 0.1312
\eea
The value of $\Phi(0) = 0.1409$ obtained in Eq. (\ref{phi0}) via $H_{u 0}^{(3)}$ and $R_\lambda^*$ is in close agreement with the aforementioned estimate. This consistency indicates that the proposed closure of the von K\'arm\'an--Howarth equation \cite{deDivitiis2026}, the prediction of $R_\lambda^*$ established in the previous section, and the present analysis provide a robust framework for describing the link between the Kolmogorov scaling law and intermittency.

We conclude this section with the following fundamental considerations. The current approach, by leveraging non-observable bifurcation modes and their corresponding quasi-PDFs, successfully recovers the Kolmogorov scaling law, which dictates that $u/u_k \sim (u/u_k)^2 (\ell_k/\lambda_T) \sim \sqrt{R_\lambda}$. This result suggests that the non-observability of bifurcation modes constitutes the primary conceptual bridge connecting the intermittency of $\Delta u_r$ to the Kolmogorov law. While intermittency might be identified through other means, the rigorous derivation of the Kolmogorov law would remain inaccessible without invoking the principle of non-observability of these modes.

\bigskip

\section{Analytical Expressions for Velocity and Temperature Increment Probability Density Functions}

Based on the preceding analysis, this section derives the analytical structure of the PDFs $F_u(\Delta u_r)$ and $F_\theta(\Delta \vartheta)$. These distributions are obtained by combining the structural relations for $\Delta u_r$ and $\Delta \vartheta$ in terms of reduced Gaussian variables [Eq. (\ref{struct funs})] with the Frobenius--Perron equations (\ref{Frobenius-Perron}) and (\ref{gaussian}). 
In the general case, the PDFs $F_u(\Delta u_r)$ and $F_\theta(\Delta \vartheta)$ do not admit a simple closed-form expression; however, they can be expressed as the convolution of a Gaussian PDF and a modified Bessel function of the second kind, $K_0$
\bea
\ds K_0(x)  = \int_0^\infty \frac{\cos(x t)}{\sqrt{1+t^2}} \ dt
\eea
 Consistent with Eqs. (\ref{struct funs}), we consider dimensionless velocity and temperature increments characterized by unit standard deviation. To this end, we introduce the variable $s$, representing either $\Delta u_r$ or $\Delta \vartheta$ depending on the context, defined as:
\bea
\begin{array}{l@{\hspace{-0.cm}}l}
\ds s =  \frac{\Delta u_r}{\sqrt{ \langle (\Delta u_r)^2 \rangle}} = 
\frac{\zeta_u + \Psi_u \left( \chi(\zeta_{u+}^2-1)-(\zeta_{u-}^2-1)\right) }{\sqrt{1+2 \Psi_u^2(1+ \chi^2) }}, 
\\\\ 
\ds s = \frac{\Delta \vartheta}{\sqrt{ \langle (\Delta \vartheta)^2 \rangle}} = 
\frac{\zeta_\theta + \Psi_\theta (\zeta_{\theta+}^2-\zeta_{\theta-}^2)}{\sqrt{1+4 \Psi_\theta^2 }}, \end{array}
\label{struct funs_dimensionless}
\eea
By applying Eqs. (\ref{Frobenius-Perron}), (\ref{gaussian}), and (\ref{struct funs_dimensionless}), the velocity increment PDF, $F_u(s)$, is given by:
\bea
\begin{array}{l@{\hspace{-0.cm}}l}
\ds F_u(s)=\int_{-\infty}^\infty 
\frac{1}{\sqrt{2 \pi} \alpha_u} \ e^{\ds -\frac{(s-\zeta)^2} {2 \alpha_u^2}}  Q_u( \zeta) \ d \zeta, \\\\
\ds Q_u(\zeta) = \frac{1}{2\pi \sqrt{\beta_1 \beta_2}} \exp\left( \frac{\Delta \beta}{4 \beta_1 \beta_2} \left( \zeta + \Delta \beta \right) \right) K_0\left( \frac{\beta_1 + \beta_2}{4 \beta_1 \beta_2} \left| \zeta + \Delta \beta \right| \right), \\\\
\ds \alpha_u = \frac{1}{\sqrt{1+2\Psi_u^2\left(1+\chi^2 \right) }}, \ \
\ds \beta_1 = {\chi \Psi_u} \ \alpha_u, \ \
\ds \beta_2 = {\Psi_u} \alpha_u, \ \ \Delta \beta=\beta_1 - \beta_2
\end{array}
\label{conv u}
\eea
This function exhibits a pronounced asymmetry, with a corresponding skewness value arising from the turbulent energy cascade. Conversely, the temperature increment PDF remains even due to local isotropy:
\bea
\begin{array}{l@{\hspace{-0.cm}}l}
\ds F_\theta(s)=\int_{-\infty}^\infty 
\frac{1}{\sqrt{2 \pi} \alpha_\theta} \ e^{\ds -\frac{(s-\zeta)^2} {2 \alpha_\theta^2}}  Q_\theta( \zeta) \ d \zeta, \\\\
\ds Q_\theta(\zeta) = \frac{1}{2\pi \beta} K_0\left( \frac{1}{2 \beta} \left| \zeta \right| \right), \\\\
\ds \alpha_\theta = \frac{1}{\sqrt{1+4\Psi_\theta^2 }}, \ \
\ds \beta =  \Psi_\theta \ \alpha_\theta, 
\end{array}
\label{conv t}
\eea
For the detailed mathematical derivation showing that the convolution of a Gaussian kernel with $K_0$ arises from the distribution of second-order moment statistics, the reader is referred to the seminal work of Wishart and Bartlett \cite{Wishart1932}.
The convolution structure given by Eqs. (\ref{conv u}) and (\ref{conv t}) between Gaussian kernel and $K_0$ is consistent with the general framework for velocity probability density functions proposed by Castaing et al. \cite{Castaing1990} and specifically discussed for passive scalars and velocity increments in \cite{Ching1991, Kailasnath1992}.
\begin{figure}[h!]
\centering
\includegraphics[scale=0.5, angle=0]{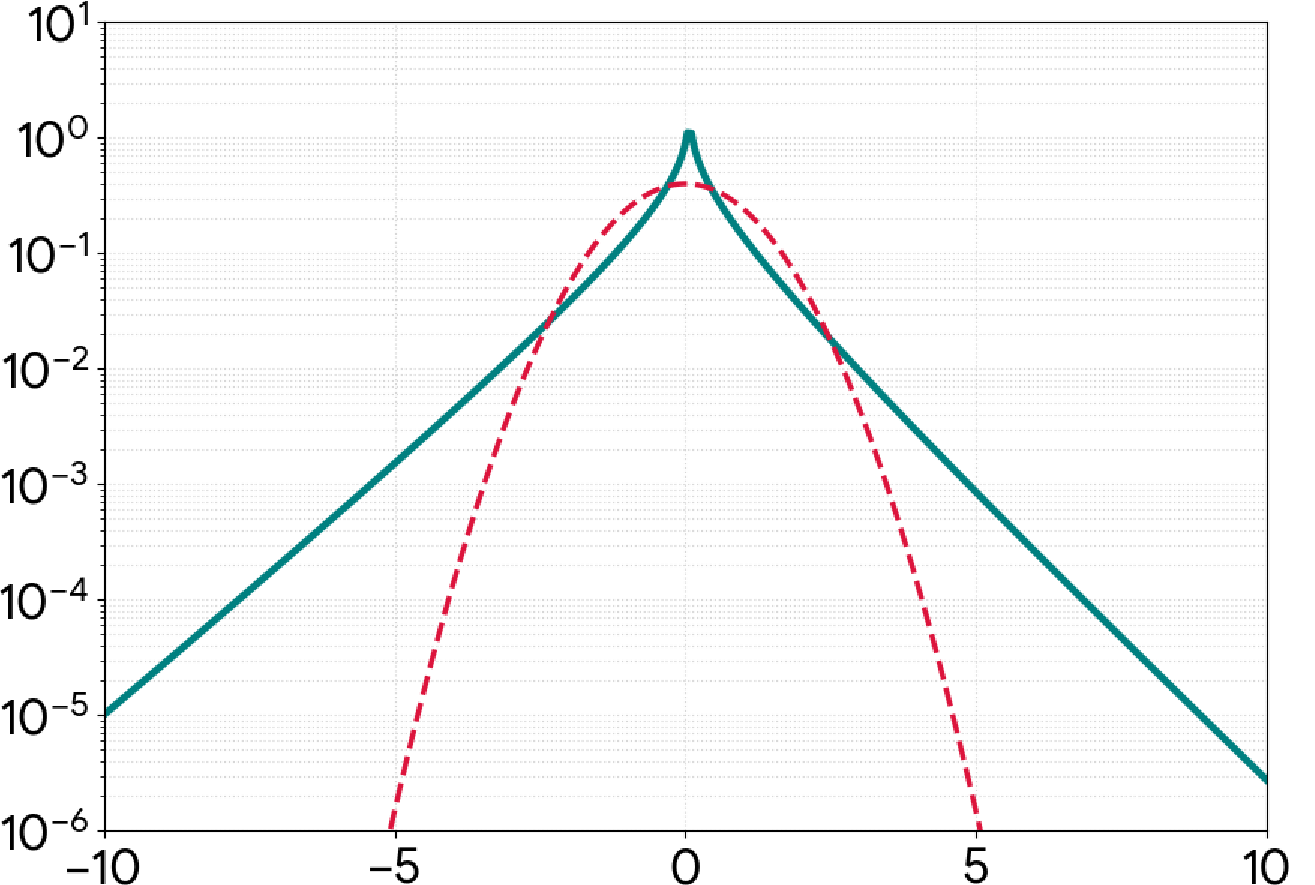} 
\caption{
Probability Density Function of $s = {\partial u_r/\partial r}/{\sqrt{\left\langle \left( \partial u_r/\partial r\right)^2\right\rangle}}$ for $R_\lambda \rightarrow \infty$ in comparison with a Gaussian PDF.
}
\label{figura p-u re infty}
\end{figure}

The convolutions (\ref{conv u}) and (\ref{conv t}) reveal that the linear-Gaussian term in Eqs. (\ref{struct funs}), which stems from fluid diffusivity (viscosity or thermal conductivity), acts as a smoothing and regularizing agent for the PDFs near the origin. In the absence of this term, a singular peak would emerge at the origin for the temperature increment PDF and near the origin for the velocity increment PDF. Thus, the Bessel function captures the effects of intermittency, while the Gaussian kernel tends to regularize this intermittency, particularly at moderate Reynolds numbers. While the skewness remains constant, the degree of intermittency increases with both the Reynolds and Peclet numbers, reaching its maximum in the limits $R_\lambda \rightarrow \infty$ and $Pe \rightarrow \infty$.
In these asymptotic limits ($R_\lambda \rightarrow \infty$ and $Pe \rightarrow \infty$), both PDFs admit exact analytical solutions. These forms are obtained as $\Psi_u \rightarrow \infty$ and $\Psi_\theta \rightarrow \infty$ using Eqs. (\ref{Frobenius-Perron}) and (\ref{gaussian}). Accordingly,  $\alpha_u \rightarrow 0$, $\alpha_\theta \rightarrow 0$, the Gaussian kernels tend to Dirac delta, and the velocity increment PDF converges to:
\bea
\ds F_u(s) = \frac{1}{2\pi \sqrt{\beta_1 \beta_2}} \exp\left( \frac{\Delta \beta}{4 \beta_1 \beta_2} \left( s + \Delta \beta \right) \right) K_0\left( \frac{\beta_1 + \beta_2}{4 \beta_1 \beta_2} \left| s + \Delta \beta \right| \right) 
\eea 
with the parameters defined as:
\bea
\ds \beta_1 = \frac{\chi_\infty}{\sqrt{2\left(1+\chi_\infty^2 \right) }}, \ 
\ds \beta_2 = \frac{1}{\sqrt{2\left(1+\chi_\infty^2 \right) }}, \  \Delta \beta=\beta_1 - \beta_2,  \ \chi_\infty \approx 0.8659,
\eea
whereas the temperature increment PDF becomes purely symmetric and is directly represented by the $K_0$ function:
\bea
\ds F_\vartheta(s) = \frac{1}{\pi} K_0(\left| s \right|)
\eea
\begin{figure}[h!]
    \centering
     \includegraphics[scale=0.5, angle=0]{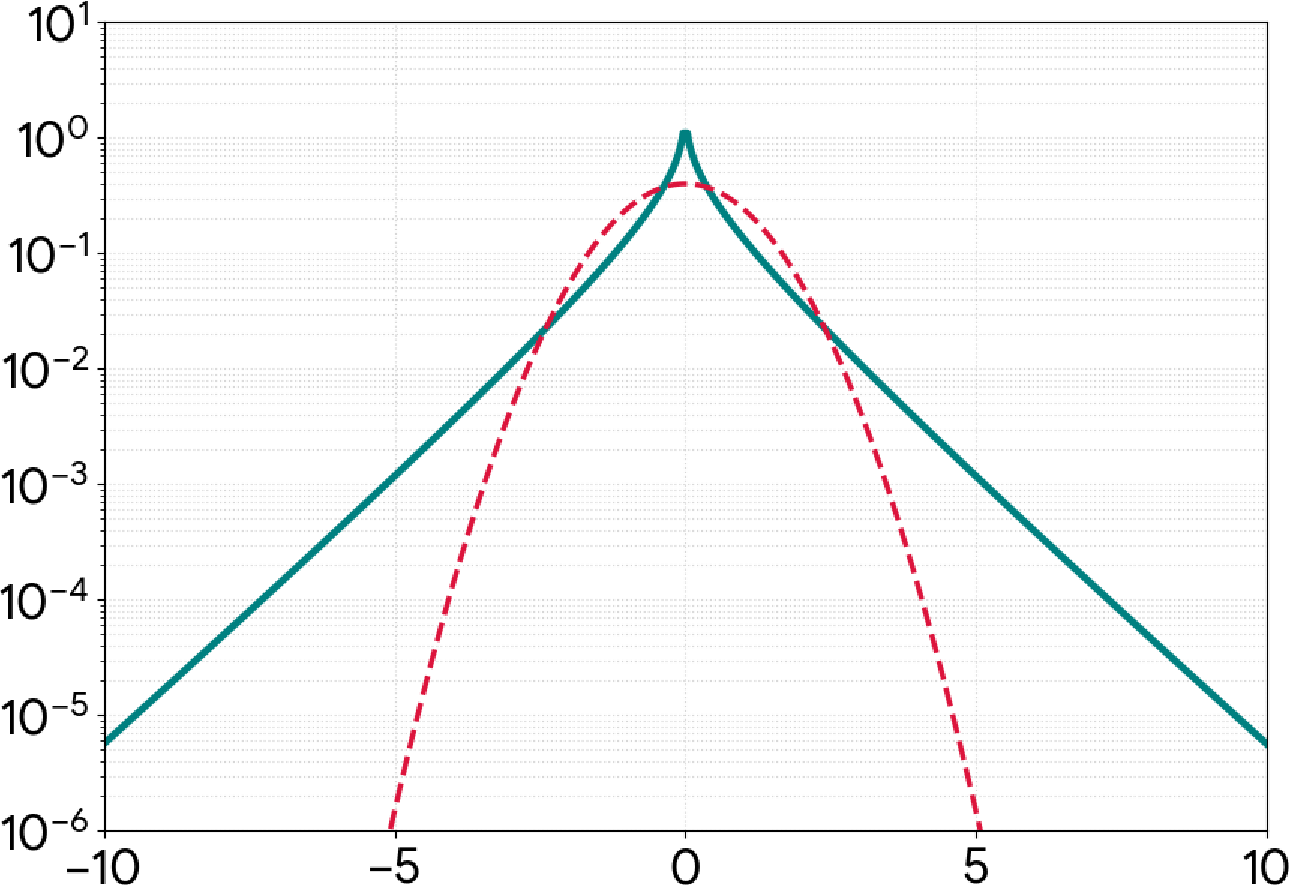} 
    \caption{Probability Density Function of $s = {\partial \vartheta/\partial r}/{\sqrt{\left\langle \left( \partial \vartheta/\partial r\right)^2\right\rangle}}$ for $Pe \rightarrow \infty$ compared with a Gaussian PDF.}
    \label{figura p-theta re infty}
\end{figure}
Figures \ref{figura p-u re infty} and \ref{figura p-theta re infty} illustrate the behavior of both PDFs under these conditions, compared to Gaussian distributions with the same standard deviation. In this regime, intermittency is maximized, as evidenced by the heavy tails dictated by the asymptotic decay of the modified Bessel function. Notably, the velocity increment PDF maintains a constant skewness of $-3/7$, which is invariant with respect to the Reynolds number. Furthermore, both PDFs exhibit a singular peak: for the temperature increment, symmetry induced by isotropy locates this singularity at the origin, whereas for the velocity increment, the singularity is shifted to $s = \beta_2 - \beta_1 \approx 0.0717$ due to the inherent asymmetry of the kinetic energy cascade. Consequently, the asymptotic decay for both PDFs follows the form $\sim \exp\left(- \left| s \right| \right) /\sqrt{\left| s \right|}$.
\begin{figure}[h!]
\centering
\includegraphics[scale=0.23, angle=0]{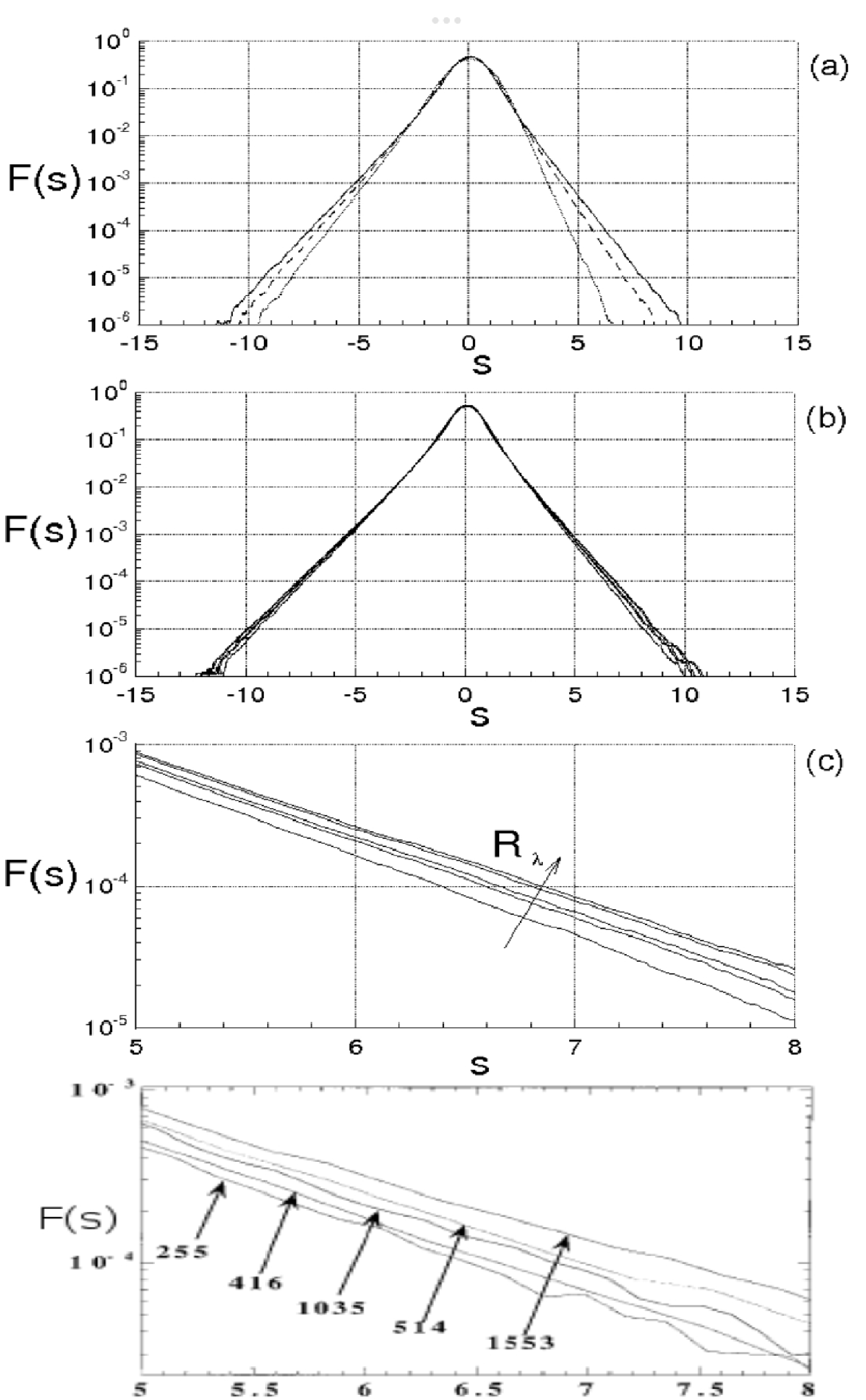} 
\caption{Comparison of the results:
Top a), b), c): PDF of longitudinal velocity derivative for various values of $R_\lambda$. a) Dotted, dash--dotted, and continuous lines represent $R_\lambda = 15, 30$, and $60$, respectively. b) and c) PDFs for $R_\lambda = 255, 416, 514, 1035$, and $1553$. Plot (c) provides a detailed view of the tails shown in (b). Bottom: Experimental data from Ref. \cite{Tabeling97}.
}
\label{figura_r5}
\end{figure}
We conclude this section by observing that the convolutions (\ref{conv u}) and (\ref{conv t}) have been derived from the analytical expressions of the previously defined velocity and temperature fluctuations, obtained in turn through the Navier--Stokes and heat equations.

\section{Results and Discussion}

In this section, we present the results obtained from the current analysis, providing a comparative evaluation with established data from the literature. 
As demonstrated in Ref. \cite{deDivitiis2011}, the statistical framework defined by Eqs. (\ref{Frobenius-Perron}) and (\ref{Tm1}) exhibits close agreement with the experimental evidence reported in Refs. \cite{Tabeling96, Tabeling97}. In those studies, the PDF of the longitudinal velocity gradient $\partial u_r/\partial r$ and its higher--order moments were measured using low--temperature helium gas within a closed cell between counter--rotating cylinders. Although those experiments involved wall--bounded flows, the measured velocity difference PDFs are remarkably consistent with the present theoretical results (Eqs. (\ref{Frobenius-Perron}) and (\ref{Tm1})). Despite a slightly non--monotonic evolution of the fourth and sixth moments ($H^{(4)}_u(0)$ and $H^{(6)}_u(0)$) observed in \cite{Tabeling96, Tabeling97}, the dimensionless statistical moments of $\partial u_r/\partial r$ follow the same trends and order of magnitude as those calculated via Eqs. (\ref{Tm1}). In particular, the PDFs derived from the present analysis successfully capture the non--Gaussian tails observed experimentally.
\begin{figure}[h!]
\centering
\includegraphics[width=100.0mm, height=70.0mm]{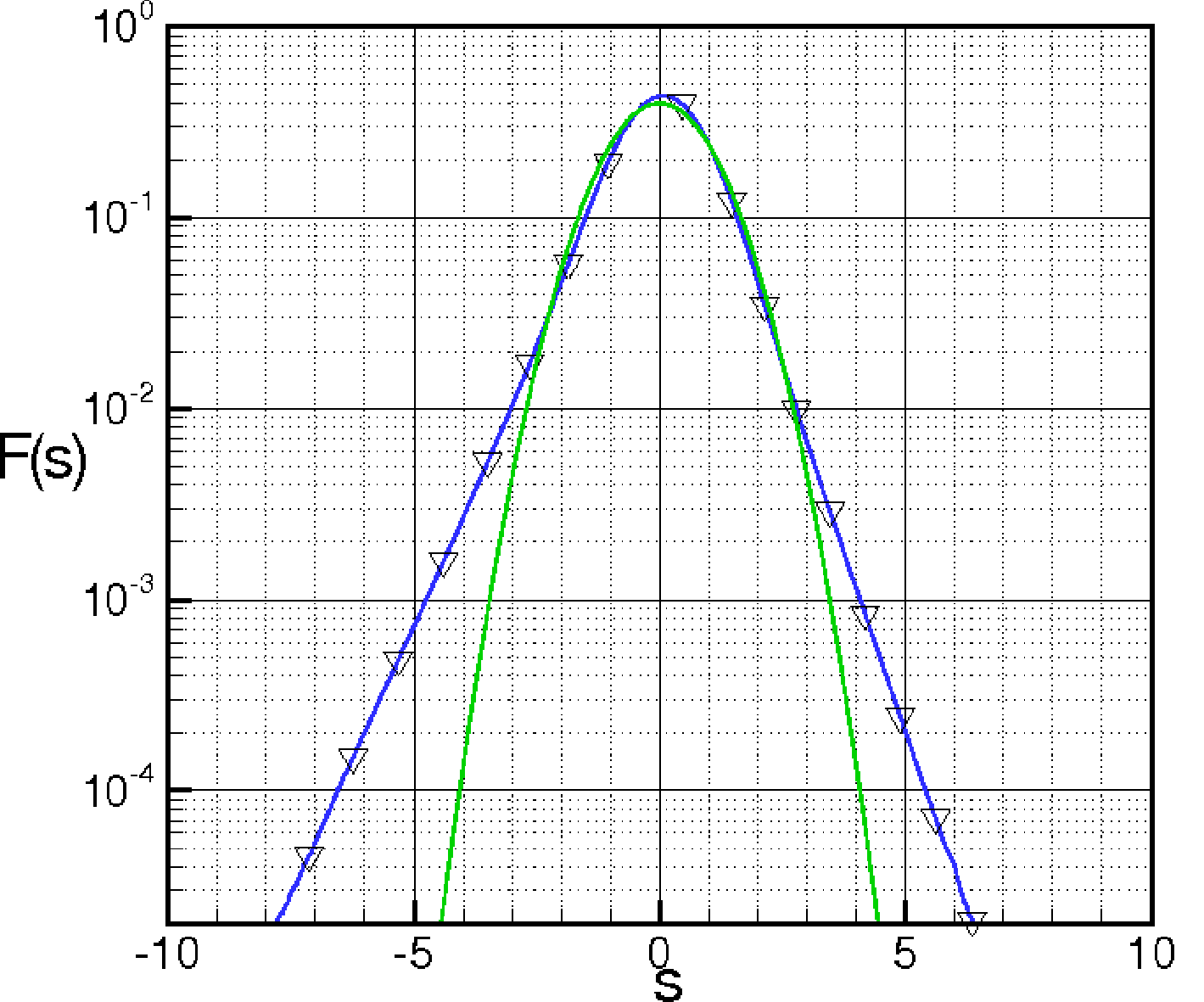} 
\caption{
Comparison of the results: PDF of longitudinal velocity differnece for $R_\lambda\simeq 1250$, $r \simeq 45 \ell_k$. Blue and green lines represent, respectively, the PDF calculated with the present analysis and the gaussian PDF with the same standard deviation. Nabla symbols are for numerical data from Ref. \cite{Yao2025}. 
}
\label{figura_r5a}
\end{figure}

In Fig. \ref{figura_r5} a), b) and c) the normalized PDFs of $\partial u_r/\partial r$, computed using Eqs. (\ref{Frobenius-Perron}) and (\ref{struct funs}), are presented in terms of the standardized variable $s$:
\bea
\ds s = \frac{\partial u_r/\partial r}{\sqrt{\left\langle \left( \partial u_r/\partial r\right)^2\right\rangle}}
\eea
such that the standard deviation is unity. The results in Fig. \ref{figura_r5}a correspond to $R_\lambda = 15, 30$, and $60$, while Figs. \ref{figura_r5}b and \ref{figura_r5}c report the PDFs for $R_\lambda$ ranging from $255$ to $1553$. Fig. \ref{figura_r5}c, in particular, highlights the PDF tails in the range $5 < s < 8$. According to the present analysis, the PDF tails broaden with the Reynolds number in the interval $10 < R_\lambda < 700$, while beyond $R_\lambda > 700$, the variations become increasingly marginal. The comparison with the results of \cite{Tabeling97} demonstrates that Fig. \ref{figura_r5}c yields PDF magnitudes and average slopes in good agreement with experimental data, particularly in the far--tail region ($s > 5$). It should be noted that the slight discrepancies and non-monotonic trends in the experimental data may arise from the departure from perfect isotropy inherent in wall--bounded flows, where boundary effects can significantly influence the velocity field near the probe.
\begin{figure}[h!]
\centering
\vspace{-0.mm}
\hspace{-7.0mm}
\includegraphics[width=100.0mm, height=70.0mm]{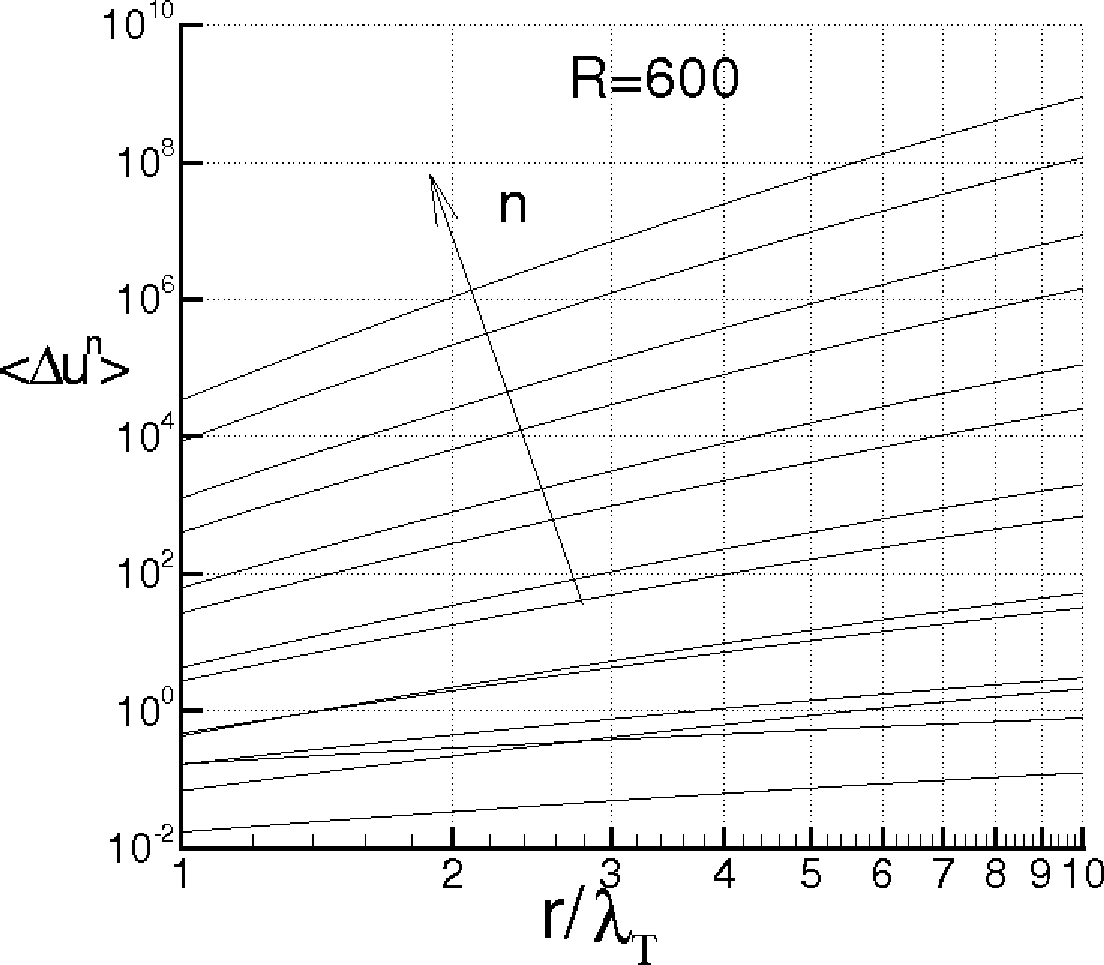} 
\hspace{7.mm}
\includegraphics[width=100.0mm, height=70.0mm]{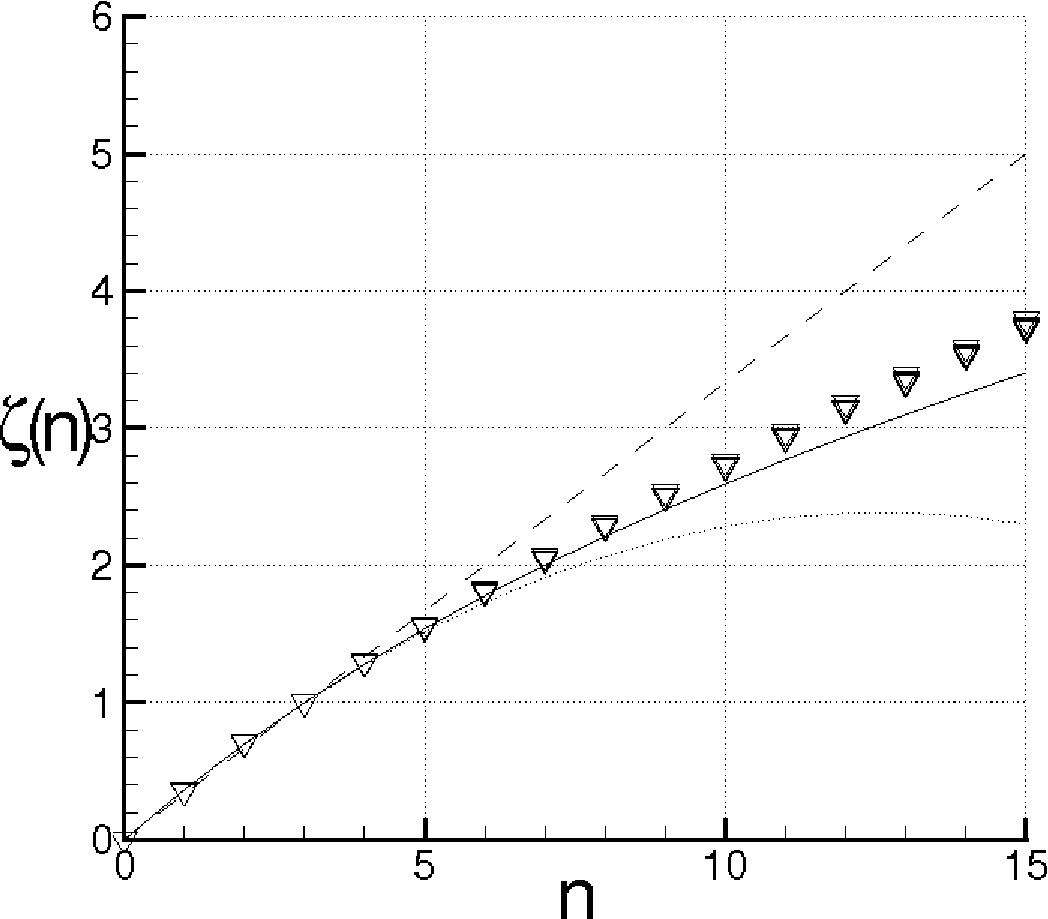}
\caption{
Top: Statistical moments of $\Delta u_r$ as a function of the separation distance for $R_\lambda=600$. Bottom: Scaling exponents of the velocity increments at different $R_\lambda$. Solid symbols denote data from the present analysis. The dashed line represents K41 theory \cite{K41}, the dotted line denotes K62 theory \cite{K62}, and the continuous line indicates the She--Leveque model \cite{She94}.
}
\label{figura_r6}
\end{figure}

{
Figure \ref{figura_r5a} presents a comparative analysis between the current theoretical results and Direct Numerical Simulation (DNS) data for forced homogeneous isotropic turbulence at $R_\lambda \simeq 1250$, as reported by Yao et al. \cite{Yao2025}. Specifically, we examine the Probability Density Function (PDF) of the normalized velocity increment:
\begin{equation}
s = \frac{\Delta u_r}{\sqrt{\langle (\Delta u_r)^2 \rangle}}
\end{equation}
which is characterized by unit standard deviation and evaluated at a separation scale of $r \simeq 45 \ell_K$ (where $\ell_K$ is the Kolmogorov length scale) in Ref.~\cite{Yao2025}.

Remarkable agreement is observed between the numerical data from Ref.~\cite{Yao2025} (triangular symbols) and the results derived from the present analytical framework (solid blue line). The DNS data points from Ref.~\cite{Yao2025} were extracted via high-resolution digital image processing. In detail, Ref.~\cite{Yao2025} reports a skewness of $-0.29$ and a kurtosis of $4.99$ for the PDF. In comparison, the present theoretical analysis---formulated through the closure of the Frobenius-Perron equation (Eq. (\ref{Frobenius-Perron})) and the structure functions (Eq. (\ref{struct funs}))---yields a distribution that fits the data with high fidelity, further demonstrating the robustness of the proposed methodology.

In particular, the values of the parameters $\chi$ and $\Psi_u$ in Eq.~(\ref{struct funs}) required to recover the aforementioned skewness and kurtosis are $\chi \simeq 0.77$ and $\Psi_u \simeq 0.64$. Note that this value of $\chi$ is slightly lower than the theoretical prediction of the present analysis, which suggests a value of approximately $0.85$. This discrepancy may be attributed to the specific forcing schemes employed in the DNS of Ref.~\cite{Yao2025} which can change the PDF with respect to unconstrained decaying turbulence. Regarding the parameter $\Psi_u$, its order of magnitude is fully consistent with the Kolmogorov scaling law, i.e., $\Psi_u \sim \sqrt{R_\lambda}$,  when $r \simeq 45 \ell_K$.
}

More in detail, Table {\ref{tab1}} shows the variations  of the dimensionless statistical moments of $\partial u_r/\partial r$ in function of $R_\lambda$ following Eq. (\ref{Tm1}). 
\begin{table}[b] 
\caption{Dimensionless statistical moments of $\partial u_r/\partial r$ at different values of the
Taylor scale Reynolds numbers. P.R. as for ''present results''.}
  \begin{center} 
  \begin{tabular}{lrrrr} 
\hline
Moment \ & $R_\lambda \approx 10$ \ & $R_\lambda=10^2$ \ & $R_\lambda=10^3$ \ & Gaussian\\[2pt] 
Order    &  P. R.                 & P. R.            &  P. R. \           & Moment \\[2pt] 
\hline \\
3        & -.428571               & -.428571         & -.428571          & 0      \\
4        &   3.96973              &  7.69530         & 8.95525           & 3      \\
5        & -7.21043               &  -11.7922        & -12.7656          & 0      \\
6        &  42.4092               &  173.992         & 228.486           & 15     \\
7        & -170.850               &  -551.972        & -667.237          & 0      \\
8        &  1035.22               &  7968.33         & 11648.2           & 105    \\
9        &  -6329.64              &  -41477.9        & -56151.4          & 0      \\
10       & 45632.5                &  617583.         & 997938.           & 945    \\
\hline
 \end{tabular}
  \end{center} 
  \vspace{-5.mm}
\label{tab1}
\end{table} 
The scaling exponents $\zeta_V(n)$ associated with the $n$-th order moments of $\Delta u_r$, defined as:
\bea
\ds \ \left\langle \left( \Delta u_r \right)^n  \right\rangle \approx A_n r^{\zeta_V(n)}, 
\eea
were determined through post--processing of the results obtained in Refs. \cite{deDivitiis2011, deDivitiis2012} using Eq. (\ref{struct funs}) via an optimization procedure. First, the statistical moments of $\Delta u_r$ are calculated as functions of $r$ (Fig. \ref{figura_r6}). Subsequently, the exponents $\zeta_V(n)$ are identified through a least-squares method applied to the following objective function:
\bea
\ds J_n(\zeta_V(n), A_n) \hspace{-1.mm} \equiv  
\int_{\hat{r}_1}^{\hat{r}_2} 
\ds ( \langle (\Delta u_r)^n \rangle - A_n r^{\zeta_V(n)} )^2 dr 
 = \min, \   n = 1, 2, \dots
\eea
where the integration limits, in the inertial range, are set to $\hat{r}_1 = 0.1$ and $\hat{r}_2$ is chosen such that $\zeta_V(3) = 1$ is satisfied. The resulting exponents are displayed in Fig. \ref{figura_r6} and compared with Kolmogorov's K41 \cite{K41} and K62 \cite{K62} theories, as well as the She--Leveque model \cite{She94}. For $n < 4$, the exponents follow the classical $\zeta_V(n) \approx n/3$ scaling. For higher orders, however, the nonlinear terms in Eq. (\ref{struct funs}) induce a distinct multiscaling behavior. The calculated exponents show excellent agreement with the She--Leveque curve, being only slightly higher for $n > 8$.

\begin{figure}[h!]
    \centering
    \hspace{-19.0mm}
    \includegraphics[width=72.0mm, height=70.0mm]{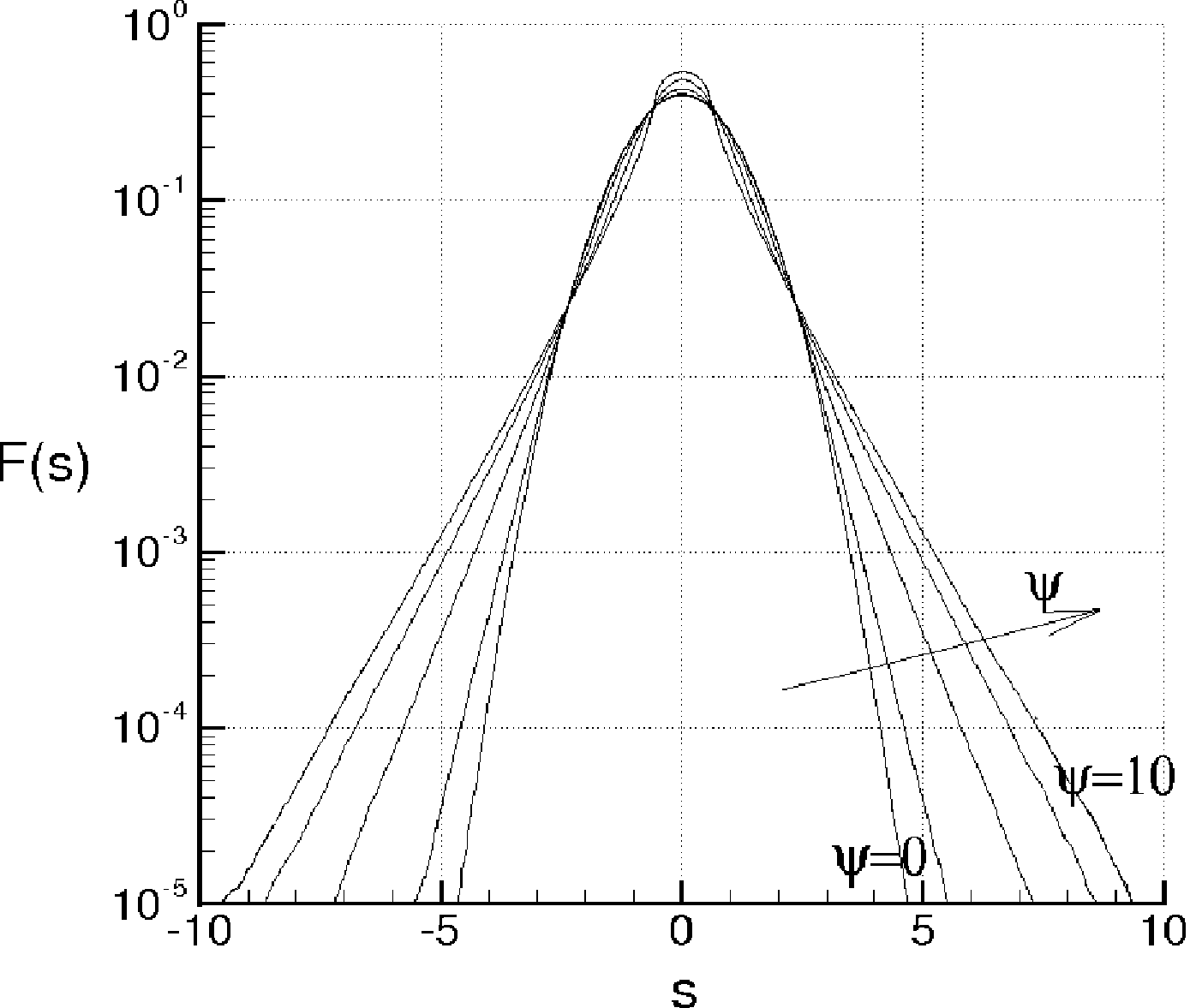}
    \hspace{4.0mm}
    \includegraphics[width=72.0mm, height=70.0mm]{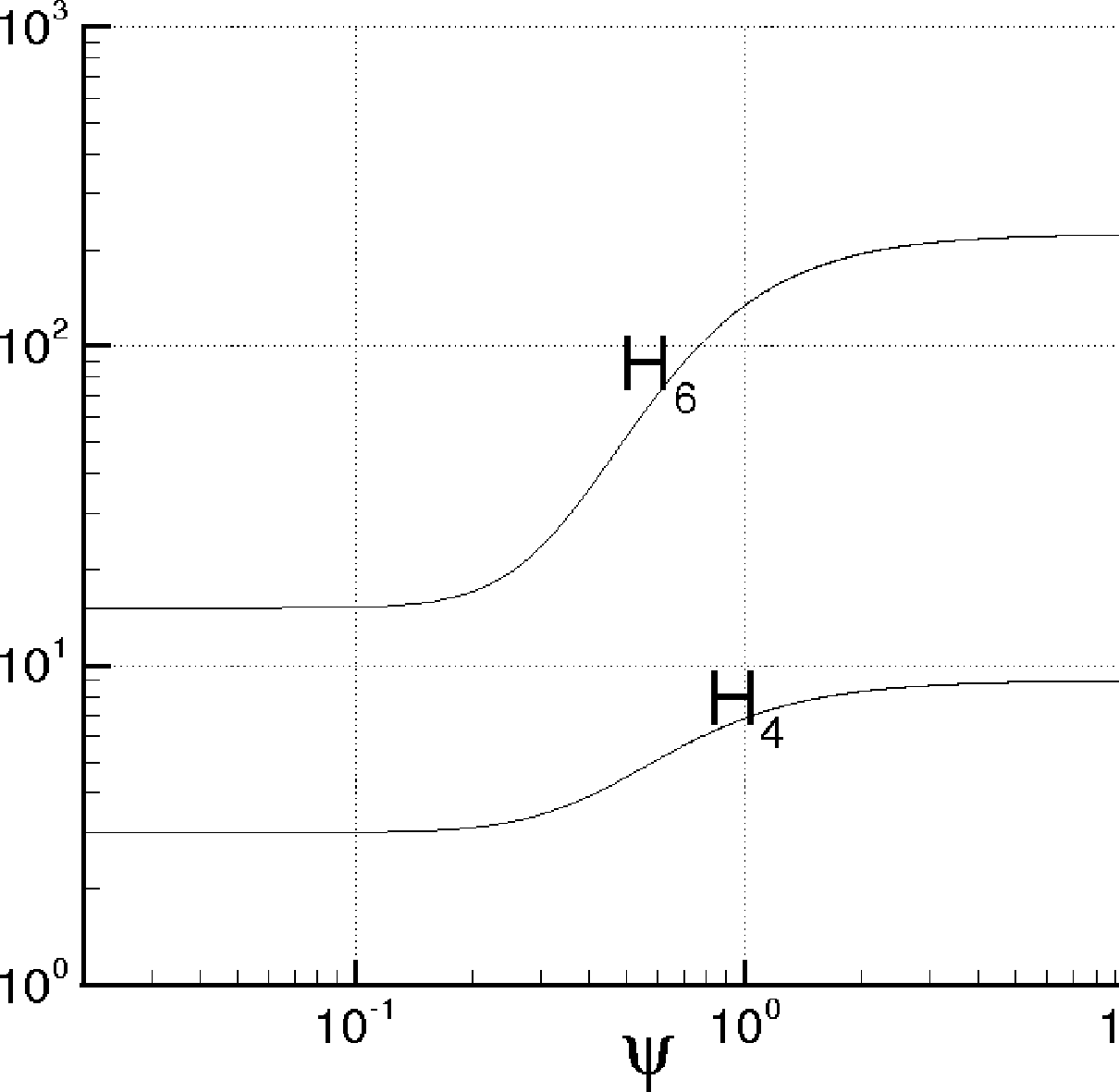}
    \caption{Left: Distribution function of the longitudinal temperature derivatives for various values of $\Psi_\theta$. Right: Dimensionless statistical moments $H_\theta^{(4)}$ and $H_\theta^{(6)}$ as functions of $\Psi_\theta$.}
    \label{figura_r7}
\end{figure}

Regarding the temperature increment statistics, Fig. \ref{figura_r7} illustrates the distribution function of $\partial \vartheta/\partial r$ in terms of the dimensionless variable:
\bea
s = \frac{\partial \vartheta /\partial r}{\sqrt{\left\langle \left( \partial  \vartheta /\partial r\right) ^2 \right\rangle} }
\eea
as computed via Eqs. (\ref{Frobenius-Perron}) and (\ref{struct funs}) for various values of $\Psi_\theta$. To quantify the intermittency of the temperature field, the flatness $H_\theta^{(4)}$ and hyperflatness $H_\theta^{(6)}$, defined as:
\bea
\ds H_\theta^{(4)} = \frac{\langle s^4 \rangle}{ \langle s^2 \rangle^2}, \quad 
\ds H_\theta^{(6)} = \frac{\langle s^6 \rangle}{ \langle s^2 \rangle^3}
\eea
are plotted in Fig. \ref{figura_r7} against $\Psi_\theta$. For $\Psi_\theta = 0$, the PDF is Gaussian ($H_\theta^{(4)} = 3$, $H_\theta^{(6)} = 15$). As $\Psi_\theta$ increases, the nonlinear contributions from $\zeta_{\theta -}$ and $\zeta_{\theta +}$ drive a significant increase in these moments, with $H_\theta^{(4)}$ and $H_\theta^{(6)}$ asymptotically approaching $9$ and $225$, respectively, as $\Psi_\theta \rightarrow \infty$.
{ These theoretical results exhibit excellent agreement with the experimental findings reported in Ref.~\cite{Sreenivasan}. In their study, the authors performed measurements characterizing, among other quantities, the kurtosis of the temperature gradient. For a Taylor-scale Reynolds number of $R_\lambda \simeq 34$, they obtained an average kurtosis value of approximately $5.5$, alongside a skewness of very small absolute magnitude. These experimental observations are in close alignment with the data presented in Fig.~\ref{figura_r7} and the results derived from Eqs.~(\ref{Frobenius-Perron})  and (\ref{struct funs}).} 

Finally, the statistical properties of the temperature dissipation rate, defined as:
\begin{equation}
 \varphi = 2 \kappa \nabla \vartheta \cdot \nabla \vartheta,
\end{equation}
are analyzed as a function of $\Psi_\theta$, with particular emphasis on its intermittent behavior. The kurtosis $K_4(\varphi)$ is estimated via Eq. (\ref{Tm1}) by invoking the assumption of local isotropy, whereby the three components of the temperature gradient $\nabla \vartheta \equiv ( \vartheta_x, \vartheta_y, \vartheta_z)$ are treated as identically distributed.

Furthermore, $\vartheta_x$, $\vartheta_y$, and $\vartheta_z$ are assumed to be statistically uncorrelated. This assumption facilitates the estimation of the kurtosis of $\varphi$ in terms of the dimensionless statistical moments of $\partial \vartheta/ \partial r$, yielding:
\begin{equation}
K_4(\varphi)=  \frac{H^{(8)}_\theta - 4 H^{(6)}_\theta +6 H^{(4)}_\theta -3}
{3 \left(  H^{(4)}_\theta-1 \right)^2} + 2
\end{equation} 
where $H_\theta^{(4)}$, $H_\theta^{(6)}$, and $H_\theta^{(8)}$ are evaluated according to Eq. (\ref{Tm1}).

\begin{figure}[h!]
    \centering
    \includegraphics[width=90mm, height=80mm]{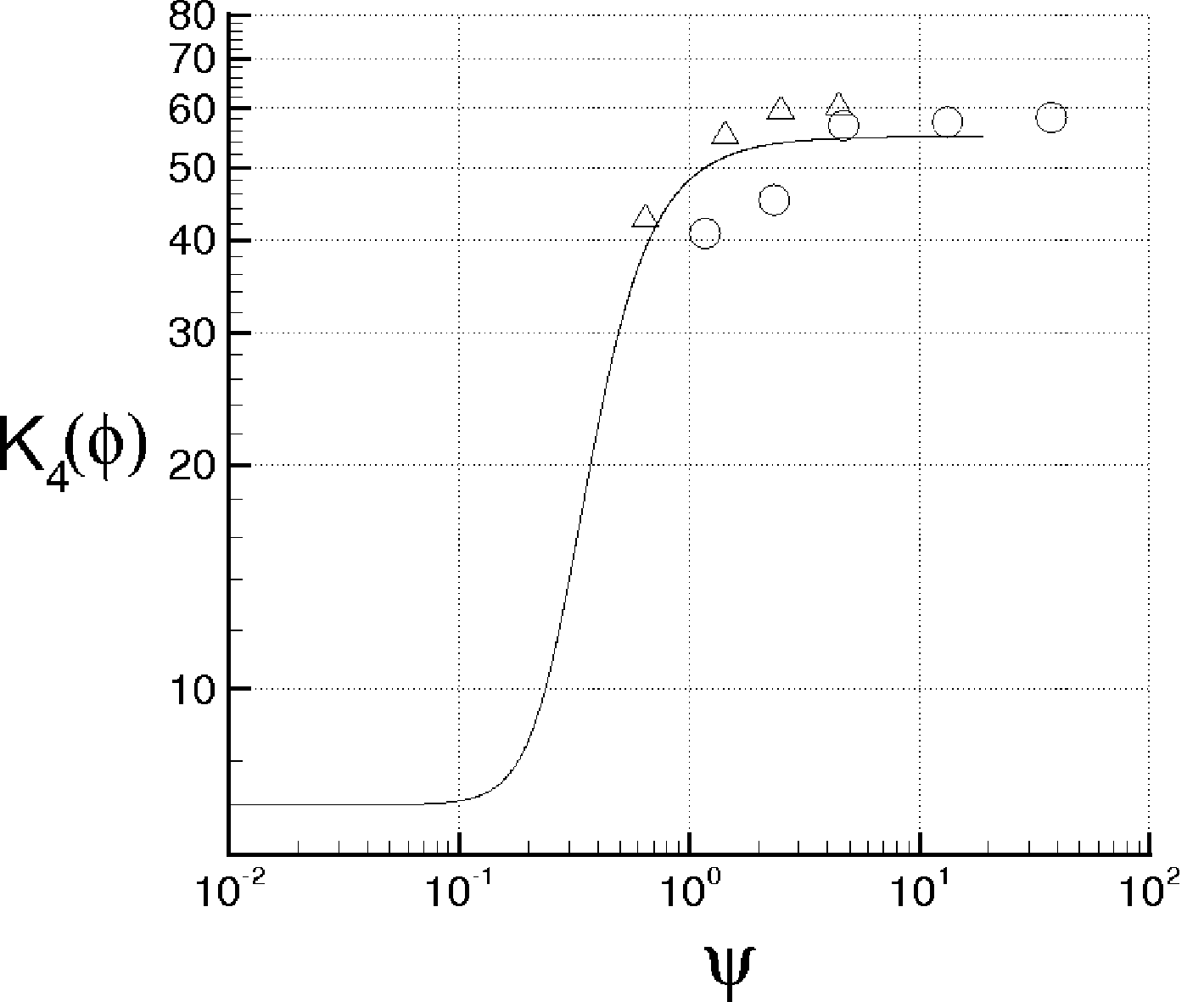}
    \caption{Comparison of the kurtosis of temperature dissipation as a function of $\Psi_\theta$. The symbols represent the results reported by \cite{Burton}.}
    \label{figura_r8}
\end{figure}

Figure \ref{figura_r8} presents $K_4(\varphi)$ as a function of $\Psi_\theta$, comparing the values derived from the present theory (solid line) with those obtained by \cite{Burton} through nonlinear large-eddy simulations (symbols). The comparison demonstrates a qualitatively robust agreement between the datasets. 

Specifically, as $\Psi_\theta \rightarrow \infty$, the current analysis predicts $K_4 \rightarrow 55$, whereas the simulations by \cite{Burton} yield a value of approximately 60. This discrepancy may be attributed to the present focus on purely isotropic turbulence  --which tends to constrain the dimensionless statistical moments of both $\partial \vartheta/ \partial r$ and $\varphi$-- as well as the simplifying assumption of statistically uncorrelated temperature gradient components.
\begin{table}[h]
\centering
\begin{tabular}{cc} 
\hline
Reference & $\Phi(0)$ \\ [2pt] 
\hline
\hline 
Present Analysis & 0.1409... \\
\cite{Tabeling96} & $\simeq$ 0.148 \\
\cite{Sreenivasan} & $\simeq$ 0.135 \\
\hline
\end{tabular}
\caption{Identification of $\Phi(0)$ through the elaboration of experimental data from \cite{Tabeling96} and \cite{Sreenivasan}, and comparison with the present analysis.}
\label{table3}
\end{table} 
Finally, it is observed that the experimental data provided by \cite{Tabeling96} and \cite{Sreenivasan} allow for the identification of $\Phi(0)$. Table \ref{table3} reports a comparison between the value of $\Phi(0)$ calculated via the proposed theory and those derived from the experimental results of \cite{Tabeling96} and \cite{Sreenivasan}. As shown, the value of $\Phi(0)$ computed using Eq. (\ref{phi0}) is in excellent agreement with the values obtained from the processing of the aforementioned experimental data. 

{
It is important to emphasize that the results of the present statistical analysis, specifically regarding the structure functions and statistical moments, are strictly valid only under the assumption of isotropy in the mean fields. In the absence of isotropy, the Central Limit Theorem (CLT) cannot, in general, be applied to the linear and quadratic forms discussed in the previous sections concerning velocity and temperature decomposition--particularly with respect to the quadratic terms. Consequently, under anisotropic conditions, the non-dimensional statistical moments could potentially exceed the calculated values, leading to levels of skewness and intermittency significantly higher than those reported in this study.
}

\bigskip

\section{Conclusion \label{Conclusion}}

In this work, we have elucidated the fundamental connection between the nonlinear dynamics of the Navier--Stokes equations and the statistical properties of velocity and temperature increments in homogeneous isotropic turbulence. The analysis demonstrates that while nonlinearity is the primary driver of intermittency, it cannot, in isolation, account for the emergence of the Kolmogorov scaling laws.

The theoretical framework is firmly rooted in the proposed closure schemes for the von K\'arm\'an--Howarth and Corrsin equations. These closures not only provide a consistent dynamical description of the energy and scalar decay but also allow for the formal identification of the transition Taylor--scale Reynolds number via specific bifurcation analysis of closed von K\'arm\'an--Howarth equation.

The central contribution of this study is the introduction of the non-observability of bifurcation modes as the essential conceptual bridge between the energy cascade and the observed statistical structure of the flow. By leveraging a decomposition into these non-observable modes and characterizing them through quasi-PDFs, we demonstrate that the synergistic effect of nonlinearity and non-observability rigorously recovers the Kolmogorov scaling laws. Specifically, we show that the ratio of the velocity standard deviation to the Kolmogorov velocity scales as $R_\lambda^{1/2}$. Moreover, our analysis reveals that the bifurcation modes exhibit amplitudes whose third statistical moment scales as $R_\lambda^{-3}$.

Significantly, the non-observability of bifurcation modes not only explains the Kolmogorov scaling law and the growing intermittency with the Taylor-scale Reynolds number, but also, and more importantly, enables the determination of the internal structure functions for velocity and temperature differences and their corresponding PDFs and statistics, which reflect very well the existing experimental and numerical data in the literature.

By invoking Fisher principle of the minimum number of parameters (1922), we have derived a closed analytical form for the statistics of velocity and temperature increments. This framework successfully captures the growth of intermittency as a function of the Taylor--scale Reynolds number, providing a theoretical justification for the increasingly non--Gaussian behavior observed in fully developed turbulence.

The consistency between the predicted normalization coefficients, the skewness values obtained via Lyapunov analysis, and the established experimental benchmarks confirms the robustness of the proposed approach. Ultimately, these results suggest that the non--observability of bifurcation modes is not merely a mathematical convenience, but a necessary physical requirement to reconcile the intermittent nature of small--scale fluctuations with the classical universal scaling laws of turbulence. 
{
In anisotropic systems, orientation jumps of the Lyapunov basis would reflect the underlying broken symmetry. While the non-observability principle still holds, the quasi-PDFs would incorporate directional biases, reconciling universal scaling with local anisotropy. In such cases, the Central Limit Theorem (CLT) is generally inapplicable to the previously discussed linear and quadratic forms. Consequently, anisotropic conditions may yield non-dimensional statistical moments, skewness, and intermittency levels significantly exceeding the values reported in the present isotropic analysis.
}

\bigskip 

\section{Acknowledgments}

This work was partially supported by the Italian Ministry for the Universities 
and Scientific and Technological Research (MIUR).

\bigskip

\end{document}